\documentclass[fleqn,12pt,twoside]{article}
\usepackage{espcrc1}
\usepackage{graphicx}
\usepackage{psfig}
\usepackage{epsfig}
\usepackage{dcolumn}
\usepackage{bm}
\usepackage{supertabular} 
\usepackage[figuresright]{rotating}

\newcommand{\AmS}{{\protect\the\textfont2
  A\kern-.1667em\lower.5ex\hbox{M}\kern-.125emS}}

\title{ Microscopic Description of Super Heavy Nuclei} 
\author{Y. K. Gambhir$^a$, A. Bhagwat$^a$ and M. Gupta$^{a,b}$  \\ 
$^a$ Department of Physics, I.I.T. Powai, Bombay 400076, India, \\
$^b$ Manipal Academy of Higher Education, Manipal 576119, India}
\begin{document}
\maketitle
\begin{abstract}
The results of extensive microscopic Relativistic Mean Field (RMF) 
calculations for the nuclei appearing in the $\alpha$ - decay chains of 
recently discovered superheavy 
elements with 109 $\leq Z \leq $ 118 are presented and discussed. The 
calculated ground state properties like total binding energies, $Q$ values, 
deformations, radii and densities closely agree with the corresponding 
experimental data, where available. The root mean square radii closely follow 
$A^{1/3}$ law ($A$ being the mass number) with the constant $
r_o=0.9639\pm0.0005$
fm. The double folding ($t\rho\rho$) approximation is used to calculate the 
interaction potential between the daughter and the $\alpha$, using RMF densities
along with the density dependent nucleon - nucleon interaction (M3Y). This in
turn, is employed within the WKB approximation to estimate the half lives 
without any additional parameter for $\alpha$ - decay. The half lives are highly
sensitive to the 
$Q$ values used and qualitatively agree with the corresponding experimental 
values. The use of
experimental $Q$ values in the WKB approximation improves the agreement with
the experiment, indicating that the resulting interaction potential is reliable
and can be used with confidence as the real part of the optical potential in
other scattering and reaction processes.
\end{abstract}
\maketitle
\section{Introduction}

The synthesis of the superheavy elements (SHE) is a subject that has remained 
at the forefront of nuclear physics ever since the prediction of islands of 
stability in this mass region (Z $\sim$ 114, N $\sim$ 184) in the 1960's 
\cite{nil69,mos69}. After about four decades of experimentation, it still
does not seem to be possible to reach the N = 184 closed shell with stable 
neutron rich beams such as $^{48}$Ca used at JINR/Dubna. In order to get 
closer, radioactive ion beams with neutron excess higher than $^{48}$Ca
will be required \cite{oga2} such as those that may become available with the
DRIBs facility at Dubna \cite{oga3} and elsewhere. 

We start with a short review of the current status of experimental work in 
Section II, followed by a brief description of the theoretical framework in
Section III. The results and discussion are contained in section IV. 
The conclusions are presented in the last (V) section.

\section{Status of Experimental Work}
The elements above Fermium have been created in the laboratory by completely 
fusing heavy ions. The evaporation residues (EVR's) of compound nuclei (CN) upto
about Z = 112 are produced primarily by successive alpha - 
decays to known nuclei 
enabling their unambiguous identification by the method of $\alpha$ - $\alpha$ 
correlations \cite{ghi74}. However, at Seaborgium, the production cross - 
sections
fall below one nanobarn with half - 
lives of less than a second. To access the new
region above Z=106, new methods had to be devised \cite{arm85,hof00} such as 
`cold-fusion' using Pb and Bi targets \cite{oga74}, invented by the Dubna Group 
and successfully used in the early discoveries of Bh, Hs and Mt. Together with 
increasingly sensitive detection methods it thus became possible to identify a 
single atom \cite{mun84}. Using such techniques, two isotopes of Hassium, 
$^{264,265}$Hs were created at GSI in the 1980's \cite{mun87,oga84a}, $^{267}$Hs
in the mid - 90's at Dubna  \cite{laz95} and $^{277}$Hs as a part of the Z = 
114 
chain \cite{oga01} also at Dubna. The GSI group had successfully produced the 
elements $^{270}$Ds \cite{hof01a}, $^{269,271}$Ds, $^{272}$111, 
\cite{hof98,hof96,hof95a,hof95b}, by 1998 and $^{277}$112 \cite{hof02} by the
year 2002. Additionally, two decay chains have come from LBNL confirming the 
$^{271}$Ds data from GSI using the same $^{208}$Pb($^{64}$Ni,{\it n}) reaction 
\cite{gin03} but with gas-filled detectors to enhance EVR collection. In the 
last couple of years new results have become available from RIKEN for 
$^{271}$110 \cite{mor04,mor05}, $^{272}$111 \cite{mor03} and for $^{278}$113 
\cite{mor07}. With extensive activity underway at this facility, the current 
experimental program includes the on - going cold fusion synthesis of recently 
completed experiments for $^{277}$112 and future plans for $^{283}$114 
\cite{mor06}. Thus, cold fusion reactions usually involving the evaporation of 
one neutron, have successfully been used to synthesize elements up to Z=112. 
However the maximum production cross-section for this element (Z = 112) is 
achieved at an excitation energy of only $\sim$ 10 MeV. If one were to 
extrapolate, to reach the heavier elements the excitation energy at the Coulomb 
barrier may well drop below zero (see \cite{arm01} for an exhaustive 
discussion). Clearly, as one approaches the predicted island of stability at 
N=184 (assuming the correctness of this postulate), a higher excitation energy 
is required. This is achievable through greater neutron excess in both the 
target and the projectiles as with `hot fusion' reactions. The Dubna group have 
successfully used actinide targets with a $^{48}$Ca beam to fuse Z$\geq $114, 
where the compound nucleus is formed following the evaporation of three or 
more neutrons. Elements $^{286,287,288,289}$114, $^{287,288}$115  and 
$^{290,291,293}$116 have thus been produced at higher excitation energies and 
different cross-sections \cite{oga99a,oga99b,oga03,oga04}. Preliminary 
results for $^{294}$118 from a single event have also been cited \cite{oga2a} 
with further experimentation in progress. As a final observation, even with hot 
fusion reactions, EVR's may not pass through a region greater than N=174-177 
\cite{ogazz}. Though this is closer by about 6-8 neutrons to the N=184 region, 
it may not be sufficient. It is hoped that with the dawn of the new era of 
radioactive ion beams the `sea of instability' may be breached. 

From the above discussion, it is clear that above Z=112, especially in reactions
involving neutron rich beams of the rare $^{48}$Ca isotope, $\alpha$-decay 
chains do not end with known nuclei making the method of $\alpha-\alpha$ 
correlations insufficient for the unambiguous identification of isotopes. That 
these chains would end in unknown regions is easily understood when recognizing 
that with each successive $\alpha$ (or $\beta$-decay), daughter nuclei move away
from the regions of stability and spherical closed shells towards instability. 
It is reasonable to expect spontaneous fission (SF) to become the preferred 
decay mode over others. SF could then be considered to be a reliable signature 
for the formation of 
extremely heavy nuclei \cite{og13}. In such cases, and ideally suited to single 
atom production, chemical methods provide for excellent assignment criteria if 
the half-lives are within the range of detection i.e. around a few seconds, with
cross-sections less than 100 pb following genetically established decays 
\cite{tue02}. Generally, cold fusion experiments produce nuclei with shorter 
half-lives than those synthesized by hot fusion.  
Hot fusion reactions are thus more suitable for chemical studies. 
Analogies may be drawn between the chemical properties of 
compounds of unknown Z with those of the chemical type of known elements 
allowing for specific assignments \cite{bar92}. By adopting quick and sensitive 
techniques for separation, it has been possible to study Rf, Db, Sg and Bh
(see \cite{schRE} for a complete review of chemistry). Furthermore, 
as $\alpha$-decay may be considered to be the signature
mechanism for the unique identification of each nuclide, genetic
correlations are unambiguous. The 
detection of an $\alpha$ or a SF fragment is 
the only means for detecting a single atom following chemical separation which 
can be done with high efficiency. The method was used very effectively in 
confirming the ${\alpha}$-decay grand-daughters
for $^{277}$112 synthesized at GSI, when 
7 decay chains from Hs isotopes were unambiguously identified
after chemical separation \cite{dul02}. 
First attempts to study the chemistry of Z = 112 were 
made by fusing natural U with a $^{48}$Ca  beam to form 
$^{283}$112 \cite{yak01,yak02,yak03}. Due to various 
problems with the experimental set-up
these results are still tentative and further experiments are underway \cite{gag04}.

Where even shorter decay half-lives occur and as production cross - sections 
get smaller, still other techniques have been devised. Investigations wherein 
masses from evaporation residues (EVR's) are identified by exploiting the 
dependence of production cross-sections on excitation energy (thus defining the 
neutron yield) together with cross-bombardments in which the mass number of 
either the projectile or target is varied (thus changing the relative yield of 
resulting {\it {xn}}-evaporation channels) have been successfully carried out 
in the past. The methods find their use in both the identification of unknown 
nuclei as well as in the study of those with short SF decay half-lives 
\cite{bar92}. This is best illustrated in the case of Z = 114 and Z = 116, where
when re-visiting results from earlier experiments using more stringent methods, 
the Russian collaborations revised  their previous conclusions 
resulting in re-assignments \cite{oga04}
for $^{286,287,288,289}$114, $^{290,291,292,293}$116 \cite{ogaae}. 
These experiments have 
been carried out
 using the Dubna Gas Filled Recoil Separator in `hot fusion' reactions 
involving $^{233,238}$U, $^{242}$Pu and $^{248}$Cm targets, collectively 
described in \cite{ogaae}. Other isotopes identified and discussed therein 
include $^{267}$Rf, $^{271}$Sg, $^{275}$Hs, $^{279,281}$Ds, and $^{282-285}$112.
It is expected that existing contradictions in the case of $^{287}$114 and 
$^{283}$112 will be resolved via continuing hot fusion studies along similar 
lines in the immediate future. Even these methods have their limitations where 
the use of cross-reactions and different decay channels become increasingly 
difficult with rapidly diminishing cross-sections for higher Z. 

It is clear that the direct determination of Z requires a stringent mass 
resolution a few mass units {$\approx$} 1 \% to measure the difference between 
a {\it{xn}} and an {\it{{$\alpha$}xn}} evaporation channel for instance. The 
mass resolution should in fact be better than 1 amu at the level of 300 amu 
(${\Delta}m/m$ $<$ 0.3 \%) \cite{og10,pop03}. It is interesting that for all 
atoms heavier than U, masses have not been measured. This is due to the 
inherently small separation efficiency of elements in the range 89 $\leq$ Z 
$\leq$ 103. The identification of nuclei produced using in-flight recoil 
separators to date is based on the kinematic characteristics of the recoil 
products - the energy of the recoils and the emission angle with respect to 
beam direction. The necessity to retain the kinematic properties of the heavy 
elements being studied imposes conditions on target thickness, background 
conditions and relatively poor (factor of $\sim$ 6) suppression of reaction 
by-products \cite{ogE7}. New generation "isotope separator on-line" (ISOL) 
based machines such as `MASHA' (Mass Analyzer for Superheavy Atoms) coming up 
at JINR-Dubna will be able to directly determine the mass of separated atoms in 
the range 112 $\leq$ Z $\leq$ 120. The limitation on measurement is set by the 
shortest measurable half-life, T$_{1/2}~\sim$ 0.5 s \cite{og12}. MASHA is 
expected to be operational shortly and this will open up new avenues for 
the study of the chemical properties of the SHE.

\section{Status of Theory}
As experimental methods have advanced considerably over the last couple of
decades, several theoretical formulations have been developed and employed to 
describe the ground state properties of each of the nuclei along the $\alpha$ - 
decay chains including binding energies, Q$_{\alpha}$ values etc. In 
spherical even-even nuclei, alpha transitions are expected to go from ground 
state to ground state and hence alpha decay energies (Q-values) serve as a 
sensitive signature for local changes in stability. This is not quite so clear 
in the case of deformed doubly even nuclei where transitions may originate from 
low lying rotational levels \cite{hyd64} or odd nuclei where transitions may be 
greatly hindered and spin isomerism may occur.

Whereas observed Q-values may, in general, be measured quite accurately, 
uncertainties do exist in experimental half-lives. Cross-section limits were
already less than 12 pb for the creation of Ds and Z=111, for instance 
\cite{mun86,oga84b} and dropped to about 1 pb for the Element 112 \cite{hof98},
resulting in poor statistics offered by single-atom count-rates. 
%
Interestingly, it is seen that $\alpha$-decay half-lives are not as precise as 
one may think in microscopic calculations either: small variations in Q-values 
(to within a couple of hundred keV) 
dramatically affect theoretical estimates of half-lives.
Given this imprecision in observed half-lives, they cannot be used in 
themselves for 
experimental assignments. However, a relatively long half-life {\it in the 
presence of} a high Q$_{\alpha}$-value, serves as a strong indicator for 
elements Z $>$ 100 \cite{bar92}. Furthermore, longer T$_{\alpha}$'s for 
even-even nuclei would imply higher hindrance factors, thereby serving to 
exclude the expected ground-to-ground transitions in such cases. Since the 
experimental presence of such hindrance factors could be considered as a 
property for assignment (or its exclusion) \cite{bar92}, theoretical predictions
and comparisons will serve as a valuable guide. Finally, the qualitative trend 
of Q-values over say, neutron numbers, could serve as a good indication of 
the transition from deformation to sphericity as one traverses the 
neighborhood of N=162 shell and goes towards the spherical Z=114 region. 
It is important that such arguments be kept in mind when comparing calculations 
with experiment.

Several theoretical investigations have been carried out using the
microscopic - macroscopic (MicMac) method and the self consistent mean field
in both the relativistic and non relativistic formalisms. The primary aim 
in early studies has been to predict the combination of neutron number (N) and  
proton number (Z) 
where spherical shell closure may occur. An ``island of stability'' had been 
predicted around the hypothetical doubly magic $^{298}$114 (N=184) about 30 
years ago. More recently, nuclei in this vicinity are expected to be spherical 
or almost so with longer half-lives. Most theories do predict N=184 as being 
magic, however, there is no consensus on the location of the proton magic 
number due to 
differences in the  treatment of the large Coulomb term and the 
spin-orbit interaction. MicMac models, which assume a prior knowledge 
about the densities and single particle potentials, include the Finite Range 
Droplet Model with folded Yukawa single particle potentials (FRDM+FY) 
\cite{mol97} and the Yukawa plus Exponential model with Woods-Saxon single 
particle potentials (YPE+WS) \cite{mun03a,mun03b}, both of which confirm the 
prediction of $^{298}$114$_{184}$ as being the next spherical doubly-magic 
nucleus. 
Non relativistic microscopic models 
such as the Skyrme-Hartree-Fock-Bogoliubov method \cite{cwi99} where the 
spin-orbit term has to be manually introduced, predict Z=120 may be as probable 
as Z=114, indicating that magic shells in this region are isotope dependent 
\cite{gre95,rut97}. Such techniques tend to overestimate the splitting of 
levels due to the spin-orbit coupling which may effect predictions for shell 
closures. With the large density of single particle states which in turn 
characterizes
this mass region, the SHE serve as a sensitive probe for distinguishing between 
the various theories that attempt to predict shell structure, especially when 
these models describe stable nuclei with comparable accuracy. Also,
it has been known for some time that deformation effects are important to the 
understanding of stability in this region \cite{cwi99}. Bohr and 
Mottelson have observed that deformation may enhance stability \cite{boh75}. 

Relativistic Mean Field theories (RMF) which incorporate the spin-orbit term 
naturally, \cite{WAL.95,SER.86,PGR.86,YKG.89,YKG.90,SER.92,LAT.94,PRI.96}  generally do 
better than the non-relativistic MF models. 
RMF has been very successful in describing ground state (g.s.) properties 
for nuclei spanning the entire periodic table, thereby claiming a 
`global fit' \cite{YKG.90,PRI.96,JPM.96}. This method has also been applied 
successfully \cite{lal96,ben99,men00,ben00,lon02,zho03,GAM.03,gen03,SIL.04,GAM.05} to describe the 
ground states of these super-transactinides where 
both pairing effects and deformation play an important role. However, it is 
known that in all self-consistent models  the occurrence of a spherical proton 
(neutron) shell closure with a given Z (N) can change with varying neutron 
number N (Z). Using the RMF formalism, neutron number N=162 has been predicted 
\cite{lal96} to exhibit shell closure at around Z~=~108~-~110. This is 
consistent with the findings of \cite{sob95}. In 
addition, the calculated shell corrections (see Tab.5, page 222 of 
\cite{lal96}) peak at N = 166 for Z = 112 indicating a stability for Z = 112 
around N = 166, in agreement with experimental observations of the superheavy 
nucleus $^{277}$112.

In this work we present comprehensive and systematic calculations for 
experimentally observed chains of SHE in the region 109 $\leq$ Z $\leq$ 118. 
Binding energies, deformations, Q-values, radii and densities
are calculated for the ground states and compared with the corresponding 
experimental values where available. The calculations agree well with the 
experiment. Most of the earlier investigations have been devoted to the 
description of the g.s. properties of SHE. Here, we focus on calculating their
half lives.

The calculated RMF densities are used to derive the 
projectile-daughter interaction energy using the Double Folding (DF) model 
with the density dependent M3Y (DDM3Y) effective nucleon-nucleon interaction.
This in turn is used to calculate the $\alpha$-decay half life of the parent 
nucleus using the WKB approximation \cite{GAM.03,GAM.05}.

\subsection{Relativistic Mean Field Theory}

The Relativistic Mean Field (RMF) theory \cite{YKG.90,PRI.96}
is now established to be one of the 
most successful approach for the accurate description of nuclear properties. 
It starts with a Lagrangian describing the Dirac spinor nucleons interacting 
via exchange of mesons and the photon. The mesons considered here are: the 
isoscalar - scalar $\sigma$, isoscalar 
- vector $\omega$ and isovector - vector $\rho$ mesons. The $\sigma$ ($\omega$)
meson produces long range attraction (short range repulsion), whereas the $\rho$
meson is necessary for the isospin dependence of the nuclear properties. The 
photon, as usual, produces the Coulomb interaction. The Lagrangian 
consists of free baryon and meson terms and the interaction terms. The Euler - 
Lagrange variational principle yields the equations of motion. In the mean field
approximation, replacing the fields by their expectation values, one gets 
a set of non-linear coupled equations: \vskip 4pt 
\noindent
{\bf 1.} The Dirac equation with potential terms involving meson and 
electromagnetic (e.m.) fields describing the nucleon dynamics \vskip 4pt 
\noindent
{\bf 2.} A set of Klein-Gordon type equations with sources involving nucleonic 
currents and densities, for mesons and the photon. 

This set of equations, known as RMF equations is  to be solved self - 
consistently. The pairing correlations, essential for the description of 
open shell nuclei, are incorporated either by simple BCS prescription, or 
self consistently through the Bogoliubov transformation. The latter leads to 
the Relativistic Hartree Bogoliubov (RHB) equations. The RHB equations 
\cite{PRI.96} read:
\begin{eqnarray}
\left(\begin{array}{cc} 
h_{D} - \lambda & \hat{\Delta} \\ 
-\hat{\Delta}^* & -h_{D}^{*} +
 \lambda \end{array}\right)
\left(\begin{array}{r} U \\ V\end{array}\right)_k~=~ 
E_k\,\left(\begin{array}{r} U \\ V\end{array}\right)_k
\label{RHB} 
\end{eqnarray}
Here, $\lambda$ is the Lagrange multiplier, $E_k$ is the quasi-particle energy 
and $U_k$ and $V_k$ are the four dimensional Dirac spinors, normalized as:
\begin{eqnarray}
\int \left(U_{k}^{\dagger}U_{k'}~+~V_{k}^{\dagger}V_{k'}\right)~=~
\delta_{kk'}~;
\end{eqnarray}
$h_{D}$ is the usual RMF Dirac Hamiltonian:
\begin{eqnarray}
h_{D}~=~-\iota{\bf \alpha}\cdot{\bf \nabla}~+~\beta\left(M~+~g_{\sigma}
\sigma\right)~+~g_{\omega}\omega^{o}~+~g_{\rho}\tau_{3}\rho_{3}^{o}~+
~e\frac{1~-~\tau_{3}}{2}A^{o}
\label{DRnuc}
\end{eqnarray}
Here M is the nucleon mass, $\sigma$ is the scalar field and $\omega^o$, 
$\rho_{3}^{o}$ and $A^{o}$ are the Lorentz time like components of the 
respective meson and e.m. fields. These fields are to be determined 
self-consistently through the Klein-Gordon (KG) equations:
\begin{eqnarray}
\left\{-\nabla^2~+~m_{\sigma}^{2}\right\}\sigma &=& 
 -g_{\sigma}\rho_{s}~-~g_{2}\sigma^{2}~-~g_{3}\sigma^{3} \label{KGsig}\\
\left\{-\nabla^2~+~m_{\omega}^{2}\right\}\omega^{o} &=& 
  g_{\omega}\rho_{v}~+~g_4{\omega^o}~^3 \label{KGome}\\
\left\{-\nabla^2~+~m_{\rho}^{2}\right\}\rho_{3}^{o} &=& 
  g_{\rho}\rho_{3} \label{KGrho}\\ 
-\nabla^2 A^{o} &=& e\rho_{c} \label{KGpho}
\end{eqnarray}  
Here, $m_\sigma$ ($g_\sigma$), $m_\omega$ ($g_\omega$) and $m_\rho$ ($g_\rho$) 
are the masses (coupling constants) of $\sigma$, $\omega$ and $\rho$ fields 
respectively; $g_2$ is the coupling constant for the cubic self interaction 
terms for the $\sigma$ field; $g_3$ ($g_4$) is the coupling constant for the 
quartic self interaction term for the $\sigma$ ($\omega$) field and $e$ is the 
electronic charge. The sources (nuclear currents and densities) appearing in the
above Klein-Gordon equations involve super spinors ($U~(V)$): 
\begin{eqnarray}
\rho_{s} &=& \sum_{E_{k}~>~0}V_{k}^{\dagger}\gamma^{0}V_{k}~, \label{s1}\\
\rho_{v} &=& \sum_{E_{k}~>~0}V_{k}^{\dagger}V_{k}~, \label{s2}\\
\rho_{3} &=& \sum_{E_{k}~>~0}V_{k}^{\dagger}\tau_{3}V_{k}~, \label{s3}\\
\rho_{c} &=& \sum_{E_{k}~>~0}V_{k}^{\dagger}\frac{1~-~\tau_{3}}{2}V_{k}
\label{s4}
\end{eqnarray}
In practice, the sum in Eqs. (\ref{s1} - \ref{s4}) is taken over the positive 
energy states (no-sea approximation).

The RHB equations have two distinct parts: the self consistent 
field ($h_{D}$) that describes the long range particle-hole correlations 
and the pairing field ($\hat{\Delta}$) that accounts for the correlations 
in the particle-particle ($pp$) channel. The pairing field $\hat{\Delta}$ 
is expressed in terms  of the matrix elements of the two body nuclear 
potential $V^{pp}$ in the $pp$-channel and the pairing tensor involving 
the super-spinors ($U,V$). In the case of the constant gap, $\hat{\Delta}$ 
($\equiv~\Delta$) becomes diagonal resulting in the BCS type expressions 
for the occupation probabilities ($v^2$) \cite{YKG.90,PRI.96}: 
\begin{eqnarray}
v_{k}^{2}~=~\frac{1}{2}\left[1~-~\frac{\epsilon_{k}~-~\lambda}
{\sqrt{\left(\epsilon_{k}~-~\lambda\right)^{2}~+~\Delta^{2}}}\right]
\end{eqnarray}
$\epsilon_k$ being the energy of the single particle state $k$ and the 
Lagrange multiplier $\lambda$ (fermi energy) is to be determined
through the BCS number equation.
As a result, the RHB equations 
(Eq. \ref{RHB}) reduce to the RMF equations with a constant gap. 

A reliable and satisfactory derivation of $V^{pp}$ is not yet achieved in RMF 
\cite{PRI.96,KUC.91}. Therefore, in practice, it is 
customary to adopt a phenomenological approach while solving the RHB 
equations. Therefore, one often uses for $V^{pp}$, the finite range 
Gogny-D1S \cite{BER.84,GON.96} interaction, which is known to have the 
right pairing content and is given by:
\begin{eqnarray}
V({\bf r_{1},r_{2}})~=~\sum_{i=1,2} e^{-\left\{({\bf r_{1}~-~r_{2}})/\mu_{i}
\right\}^{2}}  \left(W_{i}~+~B_{i}P^{\sigma}~-~H_{i}P^{\tau}~-~M_{i}P^{\sigma}
P^{\tau}\right).
\label{gognyD1S}
\end{eqnarray}
where, $\mu_{i},~W_i,~B_i,~H_i~\&~M_i$ ($i$=1,2) are parameters of the 
interaction. In the case of the constant gap approximation, the required 
gap parameters are fixed so as to reproduce the corresponding Gogny 
D1S pairing energies.

Next we present and discuss the calculated ground state properties of the 
nuclei belonging to $\alpha$ - decay chains of superheavy nuclei.

\section{Results and Discussions}

\subsection{Ground State Properties}

The RMF / RHB equations are solved either using the basis expansion
technique, or in coordinate space. Explicit numerical calculations require 
the following inputs:
\\
1)  The parameters appearing in the  Lagrangian and \\ 
2)  pairing gaps (or $V^{pp}$) 

The output includes Dirac spinors, single particle (quasi-particle) energies, 
occupancies and the mesonic and e.m. fields. 
These in turn are used to obtain
total binding energies, radii, densities, deformations etc. 

A number of Lagrangian parameter sets exist in the literature. 
Here, we employ the most frequently used  parameter set NL3 \cite{LAL.97}. 
For comparison, we also 
use the NL-SV1 parameter set \cite{MMS.00} in some cases. These sets are 
listed in Tab. (1).  
In NL-SV1,  the coupling constant ($g_4$) for the quartic self - interaction 
term of the isoscalar - vector $\omega$ field is also included, and its value
is 41.010. This inclusion improves the 
equation of state of nuclear matter. 
It is seen that results obtained by using the other 
Lagrangian parameter sets (e.g. NL1 \cite{PGR.86,YKG.90}, NL-SH \cite{MMS.93}) 
exhibit similar systematics. Therefore, the inferences made and the conclusions
drawn here are expected to hold for the other parameter sets as well. 

In this work  we solve the RMF equations in the axially symmetric deformed 
oscillator 
basis. The resulting quantities (binding energies, deformation parameters, 
matter radii, etc.) are denoted by NL3 / NL-SV1. 
The required (spherical) pairing gaps 
are tuned to reproduce the 
pairing energy obtained from RHB calculations using the Gogny D1S interaction
in the pairing channel.   

\begin{table}[htb]
\begin{center}
\caption{The different sets of Lagrangian parameters commonly used in the 
RMF / RHB calculations. The masses are in MeV. All the coupling constants are
dimensionless, except $g_2$ which is expressed in terms of $fm^{-1}$.}
\vskip 6pt
\begin{tabular}{|c||c|c|c|c||c|c|c|c|c|} \hline \hline
       &  M      & $m_{\sigma}$ & $m_{\omega}$ & $m_{\rho}$ & $g_{\sigma}$& $g_{\omega}$& $g_{\rho}$& $g_{2}$& $g_{3}$ \\ \hline \hline
NL3    & 939   & 508.194      & 782.501      & 763     & 10.217      & 12.868      & 4.474     &-10.431 &-28.885  \\ \hline
NL-SV1 & 939   & 510.035      & 783          & 763       & 10.125      & 12.727      & 4.492     &-9.241  &-15.388 \\\hline \hline
\end{tabular}
\end{center}
\end{table}

We now present and discuss the calculated ground state properties. The numerical
values of the calculated ground state properties and the decay half lives are
presented in Tab. \ref{gsp_she}.

\subsubsection{Binding Energies}

\begin{figure}[htb]
\centerline{\epsfig{file=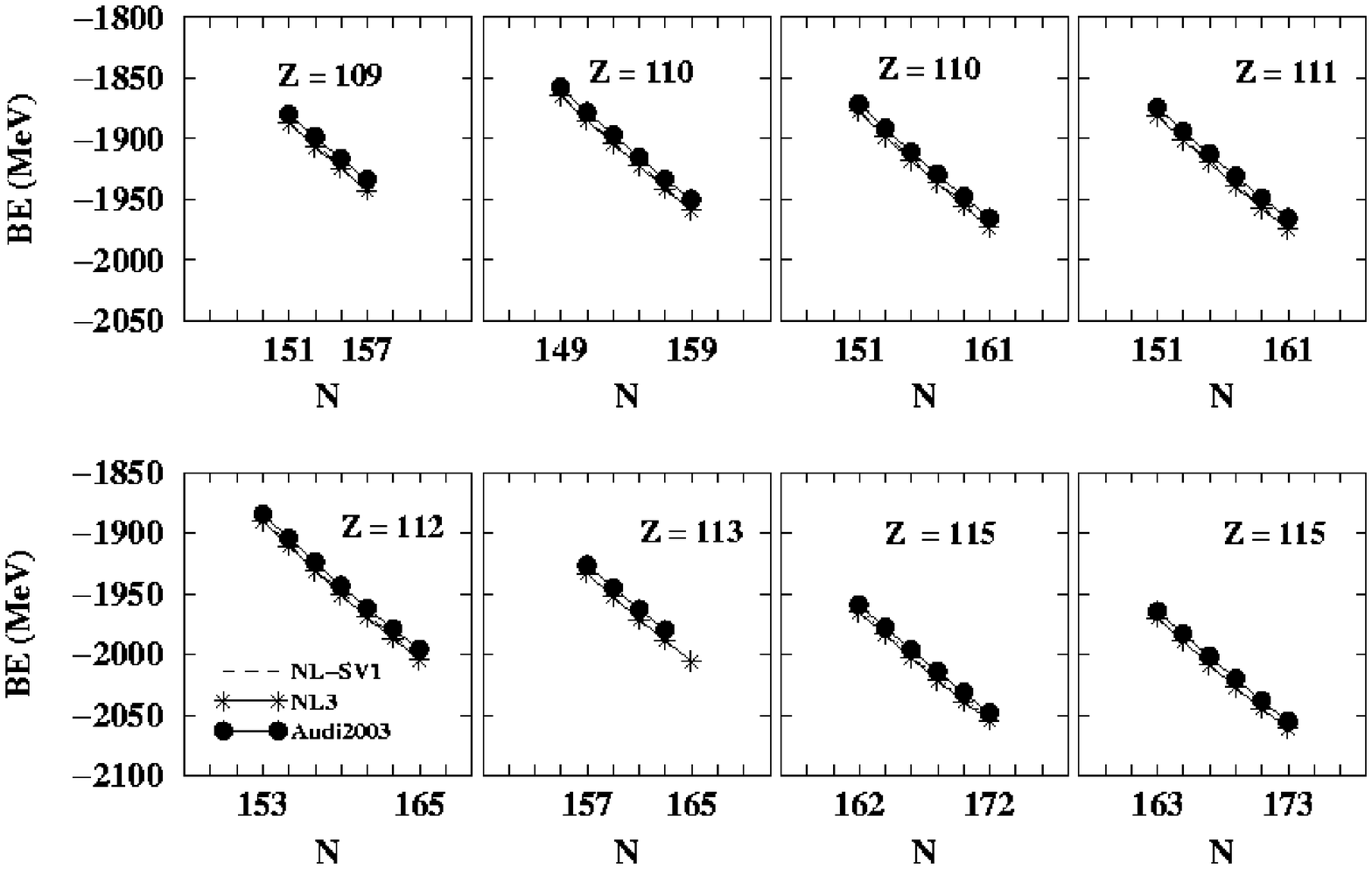,width=0.8\textwidth}}
\centerline{\epsfig{file=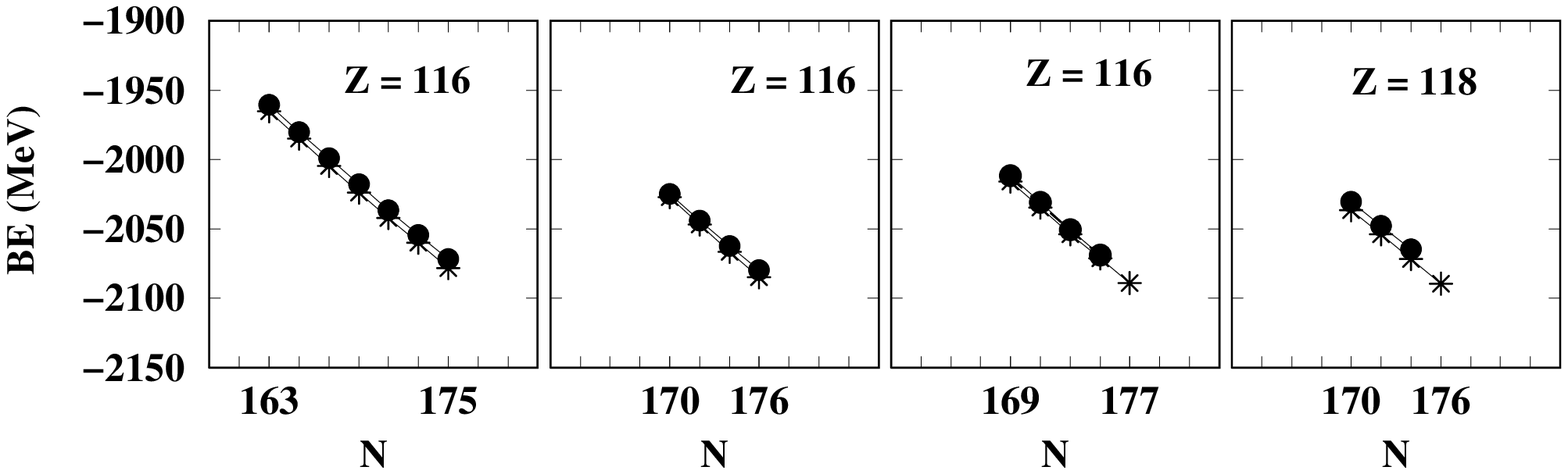,width=0.8\textwidth}} 
\caption{The calculated and the Audi2003 \cite{AUD.03} binding energies
for the nuclei belonging to $\alpha$ - decay chains of different superheavy 
nuclei.}
\label{be}
\end{figure}

The calculated and the corresponding extrapolated (Audi2003) \cite{AUD.03} binding 
energies for the nuclei appearing in some of 
the $\alpha$ - decay sequences of observed  
superheavy nuclei are presented in Fig. (\ref{be}). The DEF results obtained
using NL3 (NL-SV1) parameter set are denoted by NL3 (NL-SV1). It can be seen 
that both the calculations are in agreement with each other, and also with the
corresponding Audi2003 values. However, it should be noted that the NL3 parameter
set always yields a slightly
deeper solution in comparison with NL-SV1. Further, both
these calculations are found to yield consistently larger binding energies in 
comparison with the corresponding Audi2003 values. The differences, however, are
small (about 5 parts in 2000). 

\subsubsection{Deformations}
\begin{figure}[htb]
\centerline{\epsfig{file=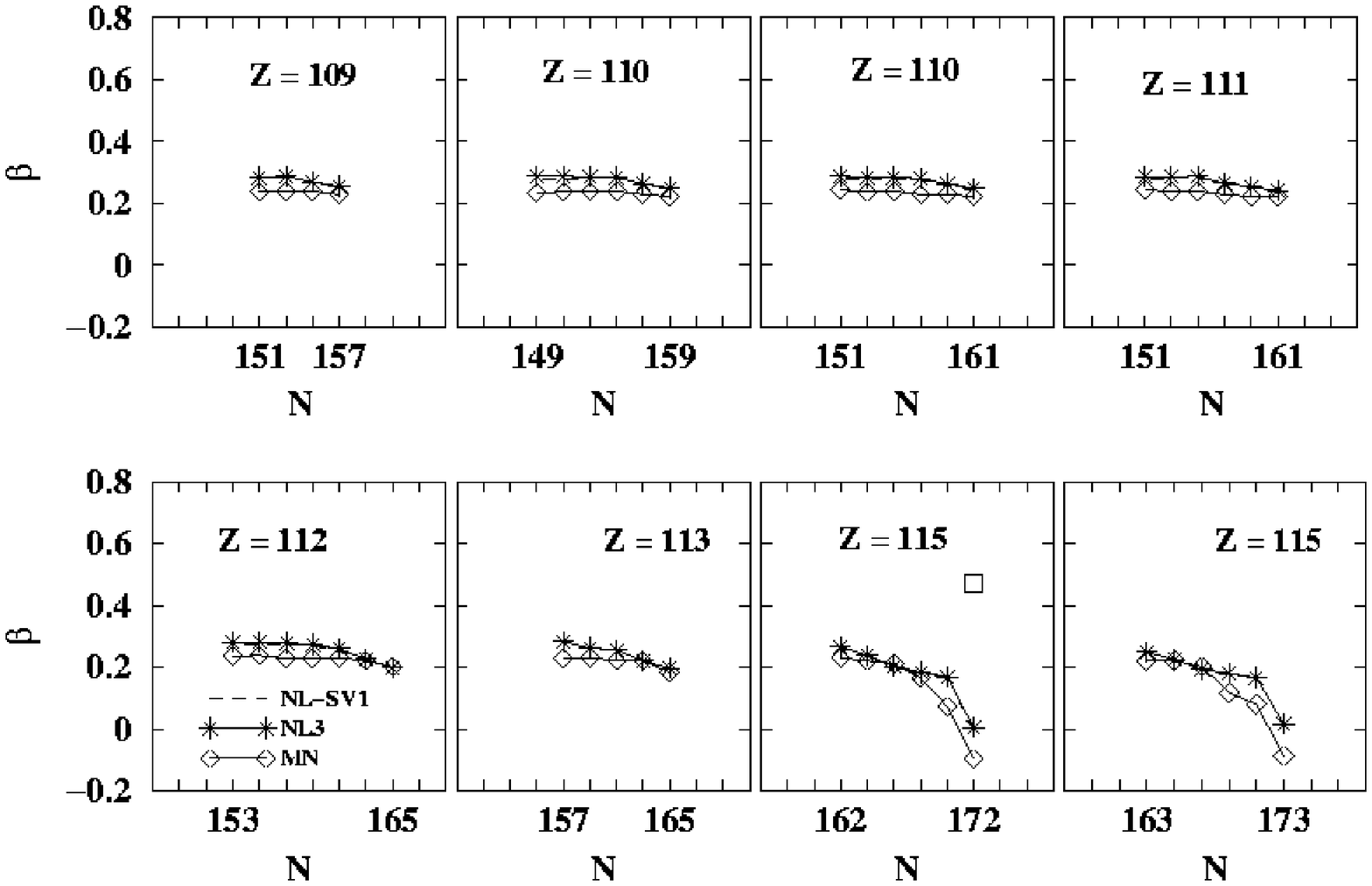,width=0.8\textwidth}}
\centerline{\epsfig{file=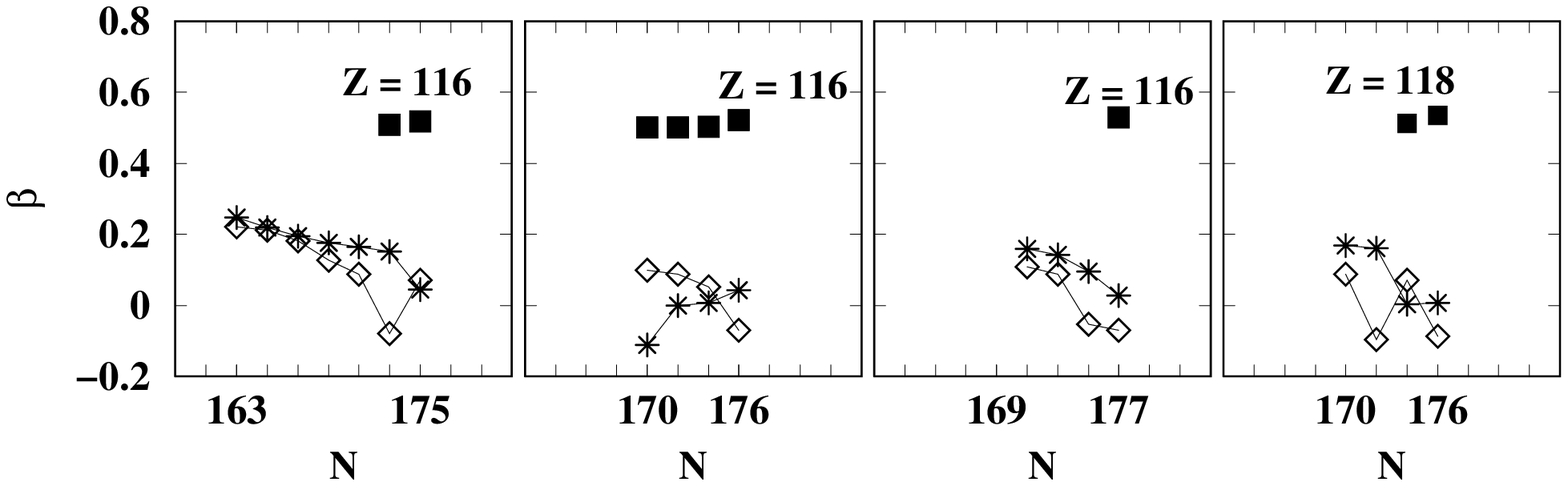,width=0.8\textwidth}}
\caption{The quadrupole deformation parameters. The corresponding Moller - Nix 
values \cite{MOL.95} are also indicated for comparison. Highly deformed
solutions are indicated by open (filled) squares for NL-SV1 (NL3) results.}
\label{beta}
\end{figure}

The quadrupole deformation parameters ($\beta$) are obtained from the calculated
point neutron ($Q_n$) and proton ($Q_p$) quadrupole moments through:
\begin{eqnarray}
Q~=~Q_n~+Q_p~=~\sqrt{\frac{16\pi}{5}}\frac{3}{4\pi}~AR_o^2\beta \nonumber
\end{eqnarray}
with $R_o~=~1.2A^{1/3}$ (fm). The calculated $\beta$ for the nuclei considered 
are shown in Fig. (\ref{beta}), along with the corresponding Moller - Nix 
(MN) values \cite{MOL.95}.  The NL3 and NL-SV1 parameter sets yield similar
deformations. These graphs reveal a number of additional interesting features.
First, most of the nuclei investigated here turn out to be prolate in shape. 
Very few of these are spherical. These findings agree with those reported
by Moller and Nix \cite{MOL.95} with some exceptions.
Further, highly deformed solutions do appear for some very heavy nuclei  
with both NL3 and NL-SV1 parameter sets. We note that  apart from these
highly deformed solutions (which happen to be the lowest), nearby
solutions also  exist with smaller values of $\beta$. 
These solutions are indicated by open / filled boxes in the respective 
figures for the 
deformation parameter $\beta$.
It has been shown  in the literature \cite{MUN.04} 
that with the incorporation of  
additional constraints, e.g. the octupole, these highly deformed 
solutions disappear. We shall ignore these solutions, and not discuss them here.

\begin{figure}[htb]
\centerline{\epsfig{file=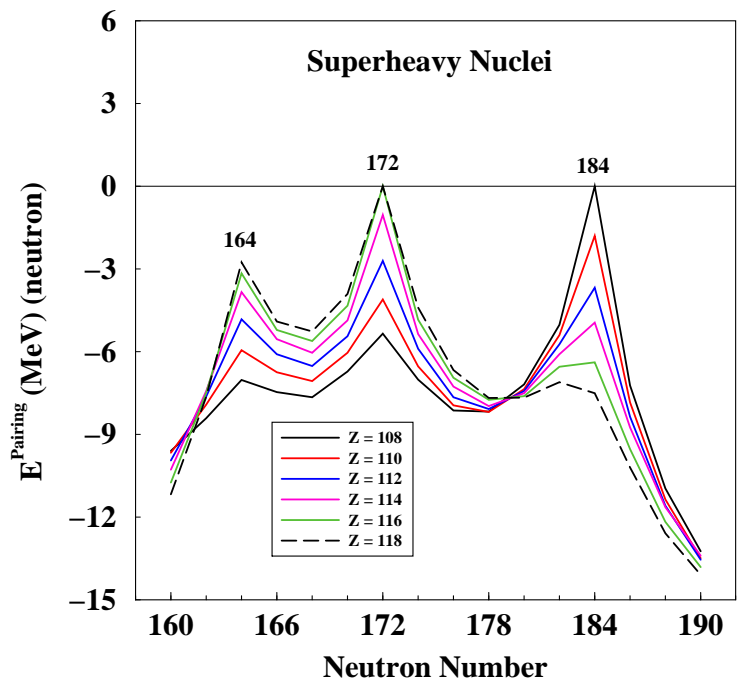,width=0.6\textwidth}}
\caption{Neutron pairing energies for even-even nuclei in the region
108 $\leq$ Z $\leq$ 118 .}
\label{pairing}
\end{figure}

\begin{figure}[htb]
\centerline{\epsfig{file=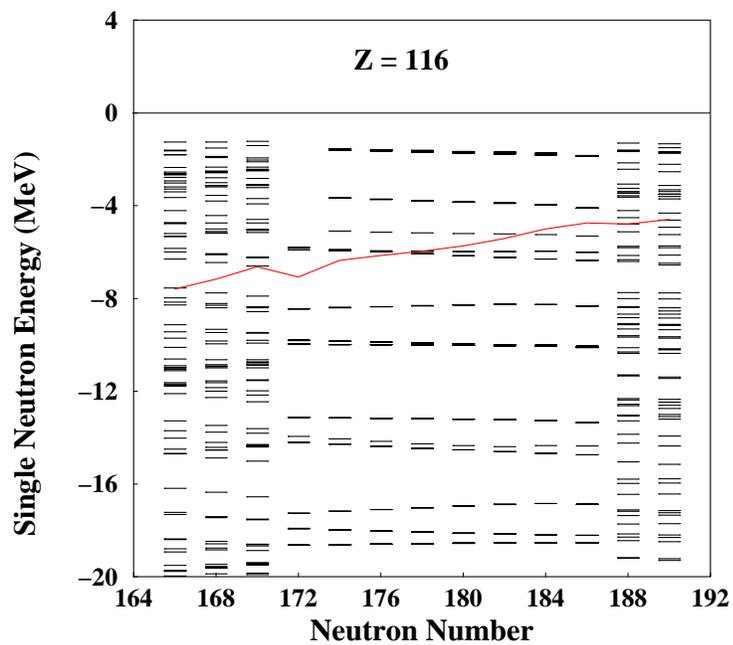,width=0.6\textwidth}}
\caption{Single particle states for Z = 116 isotopes.}
\label{116}
\end{figure}

\clearpage

\subsubsection{Pairing Energy}

One of the characteristic features of a closed shell is the existence of a sharp
maximum in the pairing energy curve. Here, we investigate the calculated 
D1S pairing energies for even-even nuclei in the 
isotopic chains of Z = 108 to Z = 118. For this 
purpose, the calculated neutron pairing energies corresponding to  
spherical solutions are shown in Fig. 
(\ref{pairing}). This figure reveals that the pairing energy for Z = 108 
at neutron number (N) 184 is zero, 
indicating that this neutron number may indicate shell closure for Z = 108. As 
$Z$ increases, the pairing energy for this neutron number also increases,
while the position of the peak is unchanged. Beyond Z = 116, however, the 
peak disappears, which may indicate loss of `magicity'. A similar `peak' 
is observed at neutron numbers 172 and 164, where, 
the peak exists for all the elements considered. Further, for 
Z = 116 and 118, the 
pairing energy turns out to be zero. For elements lighter than Z=116, the 
pairing energy is deeper (i.e. negative) and turns out to be  
deepest at  Z = 108. This 
analysis, therefore, indicates that the existence of such  `magicity' 
depends  on the  {\it combination} of both proton and neutron numbers 
rather than on either one alone.
This conclusion is consistent with the findings of \cite{lal96}.

\subsubsection{Single Particle States}
Magicity can be associated with the existence of a large gap between the 
last occupied
state and the first unoccupied state. To examine this in detail, we 
plot the calculated (NL3) single particle states for isotopes of element 116 
in Fig. (\ref{116}). The fermi levels are joined by a continuous solid line. 
A  gap is evident at neutron number 172 ($^{288}$116). This 
observation is consistent with the plot of pairing energies 
where,  at neutron number 172, the pairing energy has a sharp maximum  around 
0 MeV. Furthermore, the isotopes of Z = 116 turn out to be spherical or nearly
so, in the neighborhood of neutron number 172 ($^{288}$116). This 
fact lends support to the conclusion that for the Z=116 isotopes,
shell closure may be indicated at neutron 
number 172.

\begin{figure}[htb]
\centerline{\epsfig{file=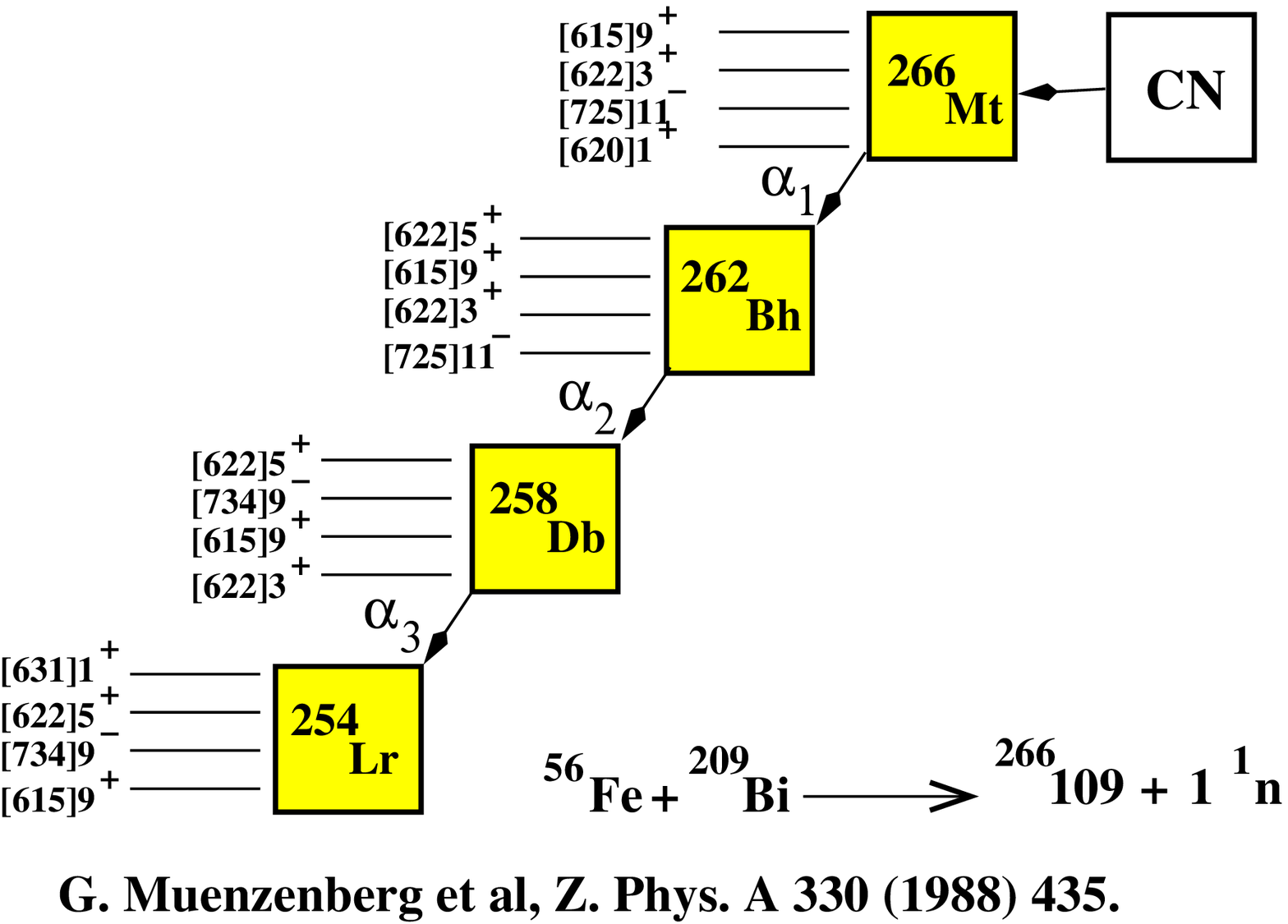,width=0.47\textwidth}}
\vspace{3cm}
\centerline{\epsfig{file=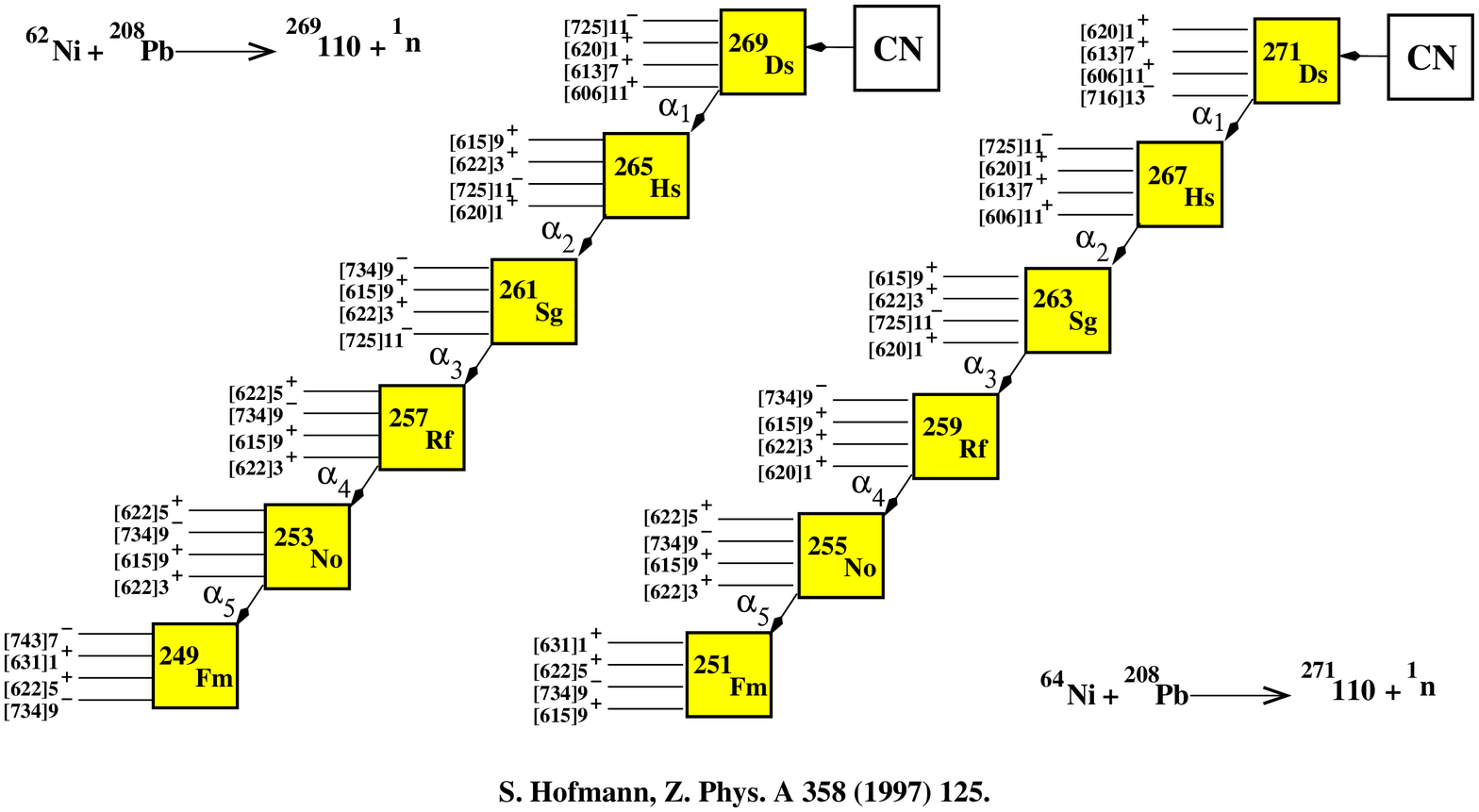,width=0.9\textwidth}}
\caption{Single particle level structure near the Fermi level}
\label{schem1}
\end{figure}

\begin{figure}[htb]
\centerline{\epsfig{file=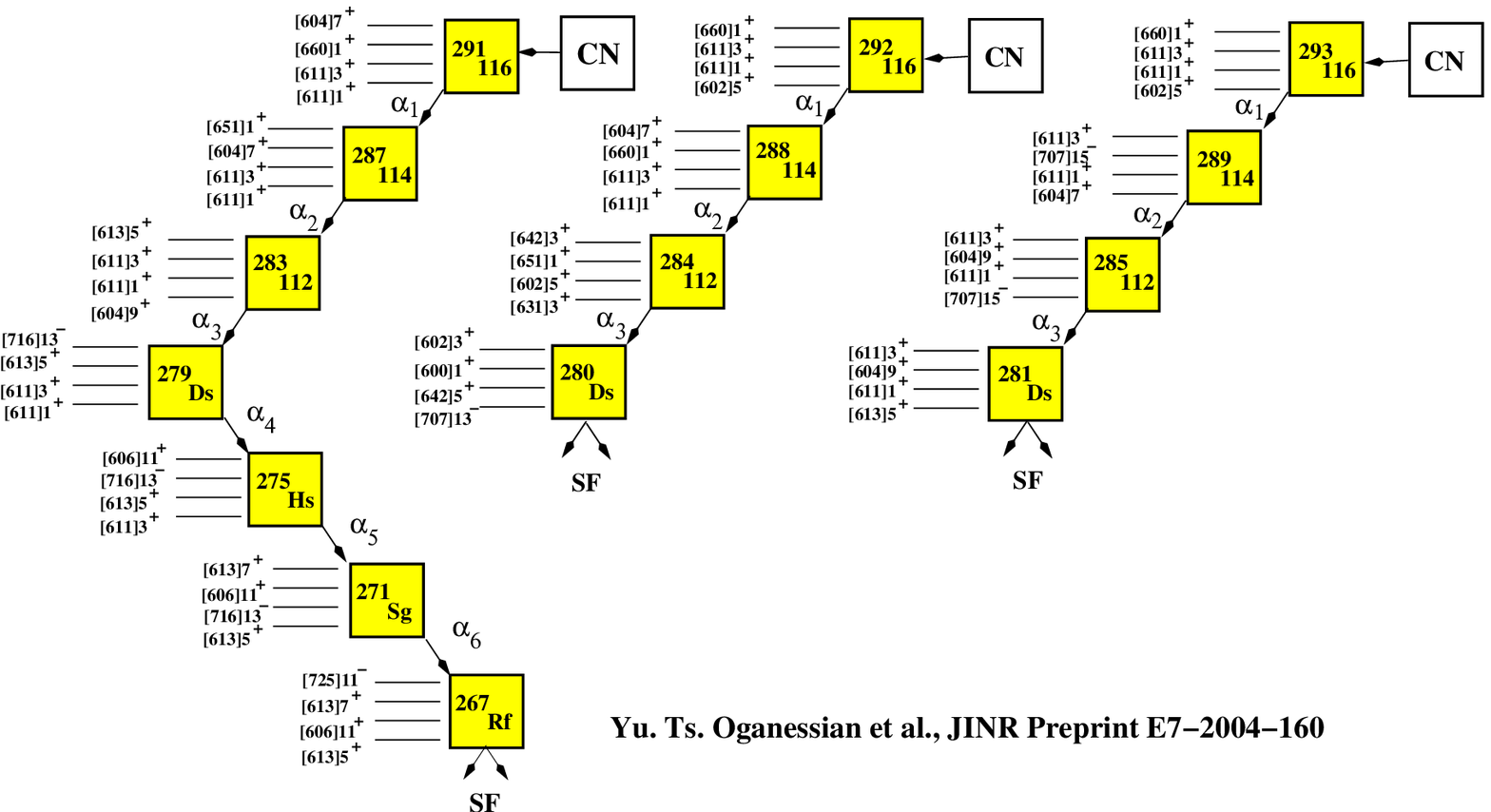,width=1.0\textwidth}}
\centerline{\epsfig{file=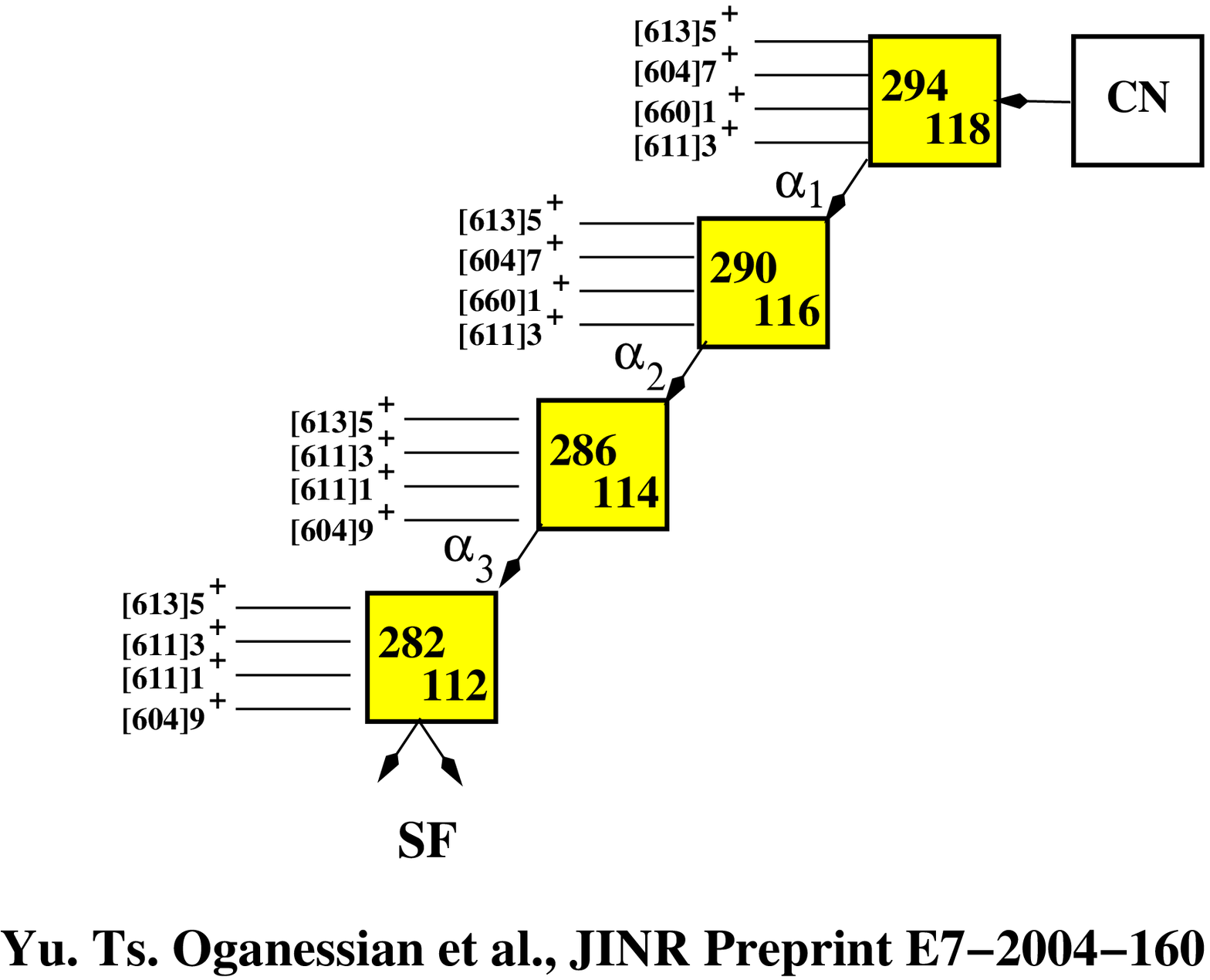,width=0.5\textwidth}}
\caption{Single particle level structure near the Fermi level}
\label{schem4}
\end{figure}

To convey the structure of the single particle levels near the Fermi surface,
we display the calculated (NL3) levels in Figs. (\ref{schem1}-\ref{schem4}) 
for the nuclei appearing in the observed
$\alpha$ - decay chains of SHE. 

\begin{small}
\begin{center}
\tablefirsthead{%
\hline
     &     &\multicolumn{2}{|c|}{BE/A (MeV)}& \multicolumn{2}{|c|}{$\beta$} &$r_o$& \multicolumn{2}{|c|}{$Q_\alpha$(MeV)}
            &\multicolumn{3}{|c|}{$\log_{10}T_\alpha$(s)} \\ \cline{3-4} \cline{5-6} \cline{8-9} \cline{10-12} 
  A  &  Z  &   NL3  &  Audi03 &   NL3  &   MN   &  (fm) &   NL3  &  Expt. & NL3    & Ex+WKB &  Expt. \\\hline }
\tablehead{%
   \hline 
   \multicolumn{12}{|l|}{\small\sl continued from previous page}\\
   \hline
     &     &\multicolumn{2}{|c|}{BE/A (MeV)}& \multicolumn{2}{|c|}{$\beta$} &$r_o$& \multicolumn{2}{|c|}{$Q_\alpha$(MeV)}
            &\multicolumn{3}{|c|}{$\log_{10}T_\alpha$(s)} \\ \cline{3-4} \cline{5-6} \cline{8-9} \cline{10-12} 
  A  &  Z  &   NL3  &  Audi03 &   NL3  &   MN   &  (fm) &   NL3  &  Expt. & NL3    & Ex+WKB &  Expt. \\\hline}
    \tabletail{%
      \hline
      \multicolumn{12}{|r|}{\small\sl continued on next page}\\
      \hline}
   \tablelasttail{\hline}
\topcaption{The calculated ground state properties of SHE. 
The $Q$ values and decay half lives are also indicated. Here, Ex+WKB represents
the half lives obtained by using the experimental $Q$ values in WKB.}
\label{gsp_she}
\begin{supertabular}{|c|c|c|c|c|c|c|c|c|c|c|c|} 
 294 & 118 &  -7.11 &        &   0.01 &  -0.09 &   0.96 &  11.18 &  11.81 &  -1.67 &  -3.21 &  -2.74 \\\hline
 293 & 116 &  -7.13 &        &   0.03 &  -0.07 &   0.96 &  10.63 &  10.69 &  -0.95 &  -1.12 &  -1.21 \\
 292 & 116 &  -7.13 &  -7.12 &   0.04 &  -0.07 &   0.96 &   9.91 &  10.80 &   1.16 &  -1.36 &  -1.74 \\
 291 & 116 &  -7.14 &  -7.12 &   0.04 &   0.07 &   0.96 &  11.69 &  10.89 &  -3.70 &  -1.72 &  -2.20 \\
 290 & 116 &  -7.14 &  -7.12 &   0.00 &   0.07 &   0.96 &  12.00 &  11.00 &  -4.33 &  -1.94 &  -1.82 \\\hline
 288 & 115 &  -7.15 &  -7.14 &   0.02 &  -0.09 &   0.96 &  12.40 &  10.61 &  -5.49 &  -1.23 &  -1.06 \\
 287 & 115 &  -7.16 &  -7.14 &   0.01 &  -0.10 &   0.96 &  13.05 &  10.74 &  -6.79 &  -1.58 &  -1.49 \\\hline
 289 & 114 &  -7.16 &  -7.16 &   0.10 &  -0.05 &   0.96 &  10.80 &   9.96 &  -2.10 &   0.22 &   0.41 \\
 288 & 114 &  -7.17 &  -7.16 &   0.01 &   0.05 &   0.96 &   9.46 &  10.08 &   1.88 &  -0.01 &  -0.10 \\
 287 & 114 &  -7.18 &  -7.16 &   0.15 &  -0.08 &   0.96 &  10.52 &  10.16 &  -1.36 &  -0.37 &  -0.29 \\
 286 & 114 &  -7.18 &  -7.16 &   0.16 &  -0.10 &   0.96 &  10.99 &  10.35 &  -2.56 &  -0.86 &  -0.80 \\\hline
 284 & 113 &  -7.20 &  -7.18 &   0.17 &   0.08 &   0.96 &  10.87 &  10.15 &  -2.57 &  -0.65 &  -0.32 \\
 283 & 113 &  -7.21 &  -7.18 &   0.17 &   0.07 &   0.96 &  10.44 &  10.26 &  -1.43 &  -0.94 &  -1.00 \\
 278 & 113 &  -7.21 &        &   0.20 &   0.18 &   0.96 &  11.20 &  11.85 &  -3.36 &  -4.88 &  -3.46 \\\hline
 285 & 112 &  -7.21 &  -7.20 &   0.14 &   0.09 &   0.96 &   9.34 &   9.28 &   1.41 &   1.61 &   1.46 \\
 284 & 112 &  -7.20 &  -7.20 &   0.00 &   0.09 &   0.96 &        &        &        &        &        \\
 283 & 112 &  -7.22 &  -7.20 &   0.17 &   0.09 &   0.96 &   9.95 &   9.67 &  -0.45 &   0.39 &   0.60 \\
 282 & 112 &  -7.22 &  -7.20 &   0.17 &   0.09 &   0.96 &        &        &        &        &        \\
 277 & 112 &  -7.24 &  -7.20 &   0.20 &   0.20 &   0.96 &  10.93 &  11.62 &  -3.02 &  -4.66 &  -3.55 \\\hline
 280 & 111 &  -7.24 &  -7.21 &   0.18 &   0.12 &   0.96 &   9.65 &   9.87 &   0.10 &  -0.55 &   0.56 \\
 279 & 111 &  -7.25 &  -7.22 &   0.18 &   0.16 &   0.96 &   9.71 &  10.52 &  -0.09 &  -2.34 &  -0.77 \\
 274 & 111 &  -7.26 &  -7.22 &   0.22 &   0.22 &   0.96 &  11.17 &  11.31 &  -3.96 &  -4.31 &  -2.03 \\
 272 & 111 &  -7.26 &  -7.23 &   0.24 &   0.22 &   0.96 &  10.92 &  10.98 &  -4.12 &  -3.37 &  -2.69 \\\hline
 281 & 110 &  -7.24 &  -7.23 &   0.16 &   0.11 &   0.96 &        &        &        &        &        \\
 279 & 110 &  -7.25 &  -7.23 &   0.18 &   0.13 &   0.96 &   9.22 &   9.84 &   1.06 &  -0.83 &  -0.74 \\
 273 & 110 &  -7.28 &  -7.25 &   0.23 &   0.22 &   0.96 &  10.74 &  11.25 &  -3.23 &  -4.46 &  -3.96 \\
 271 & 110 &  -7.28 &  -7.25 &   0.25 &   0.22 &   0.96 &  10.47 &  10.91 &  -2.46 &  -3.57 &  -3.21 \\
 269 & 110 &  -7.28 &  -7.25 &   0.25 &   0.22 &   0.96 &  10.55 &  11.28 &  -2.58 &  -4.36 &  -3.62 \\\hline
 276 & 109 &  -7.28 &  -7.25 &   0.19 &   0.20 &   0.96 &   8.88 &   9.85 &   1.77 &  -1.21 &  -0.14 \\
 275 & 109 &  -7.28 &  -7.26 &   0.20 &   0.21 &   0.96 &   9.30 &  10.48 &   0.43 &  -2.93 &  -2.01 \\
 270 & 109 &  -7.30 &  -7.27 &   0.26 &   0.22 &   0.96 &   9.79 &  10.18 &  -0.96 &  -2.04 &  -2.15 \\
 268 & 109 &  -7.30 &  -7.27 &   0.25 &   0.22 &   0.96 &  10.48 &  10.38 &  -3.60 &  -2.45 &  -1.14 \\
 266 & 109 &  -7.30 &  -7.27 &   0.26 &   0.23 &   0.96 &   9.82 &        &  -0.88 &        &  -2.32 \\\hline
 275 & 108 &  -7.29 &  -7.27 &   0.20 &   0.18 &   0.96 &   8.54 &   9.44 &   2.56 &  -0.36 &  -0.82 \\
 269 & 108 &  -7.32 &  -7.29 &   0.26 &   0.23 &   0.97 &   9.41 &   9.37 &  -0.18 &  -0.07 &   1.29 \\
 267 & 108 &  -7.32 &  -7.29 &   0.26 &   0.23 &   0.96 &   9.41 &  10.03 &  -0.19 &  -1.90 &  -1.13 \\
 265 & 108 &  -7.33 &  -7.29 &   0.26 &   0.23 &   0.96 &   9.44 &  10.73 &  -0.24 &  -3.68 &  -2.63 \\\hline
 272 & 107 &  -7.31 &  -7.29 &   0.22 &   0.22 &   0.96 &   9.25 &   9.15 &  -0.14 &   0.16 &   0.99 \\
 271 & 107 &  -7.32 &  -7.30 &   0.24 &   0.22 &   0.97 &   9.69 &        &  -1.46 &        &        \\
 266 & 107 &  -7.34 &  -7.31 &   0.26 &   0.23 &   0.97 &   8.94 &   9.22 &   0.95 &   0.07 &   0.39 \\
 264 & 107 &  -7.35 &  -7.32 &   0.27 &   0.23 &   0.96 &   8.71 &   9.77 &   1.16 &  -1.46 &   0.16 \\
 262 & 107 &  -7.34 &  -7.32 &   0.27 &   0.24 &   0.96 &  10.32 &  10.37 &  -2.93 &  -3.06 &  -2.38 \\\hline
 271 & 106 &  -7.32 &  -7.31 &   0.22 &   0.21 &   0.97 &   8.72 &   8.65 &   1.16 &   1.41 &   2.16 \\
 265 & 106 &  -7.36 &  -7.33 &   0.27 &   0.23 &   0.97 &   8.63 &   8.90 &   1.63 &   0.72 &   1.38 \\
 263 & 106 &  -7.36 &  -7.34 &   0.28 &   0.23 &   0.97 &   9.32 &   9.43 &  -0.47 &  -0.76 &  -0.93 \\
 261 & 106 &  -7.37 &  -7.34 &   0.28 &   0.24 &   0.97 &  10.13 &   9.62 &  -2.67 &  -1.30 &  -1.47 \\\hline
 268 & 105 &  -7.35 &  -7.33 &   0.25 &   0.22 &   0.97 &        &        &        &        &        \\ 
 267 & 105 &  -7.36 &  -7.34 &   0.27 &   0.23 &   0.97 &        &        &        &        &        \\ 
 262 & 105 &  -7.38 &  -7.35 &   0.28 &   0.23 &   0.97 &        &        &        &        &        \\
 260 & 105 &  -7.38 &  -7.36 &   0.29 &   0.24 &   0.97 &   9.63 &   9.34 &  -1.19 &  -0.86 &  -0.24 \\
 258 & 105 &  -7.39 &  -7.36 &   0.29 &   0.24 &   0.97 &   9.33 &   9.15 &  -0.77 &  -0.24 &   0.46 \\\hline
 267 & 104 &  -7.36 &  -7.34 &   0.25 &   0.22 &   0.97 &        &        &        &        &        \\
 261 & 104 &  -7.40 &  -7.37 &   0.28 &   0.23 &   0.97 &   8.24 &   8.65 &   2.23 &   0.84 &   0.67 \\
 259 & 104 &  -7.40 &  -7.38 &   0.28 &   0.24 &   0.97 &   8.70 &   9.02 &   0.77 &  -0.24 &   0.23 \\
 257 & 104 &  -7.41 &  -7.38 &   0.28 &   0.24 &   0.97 &   8.88 &   8.71 &   0.22 &   0.68 &   1.14 \\\hline
 256 & 103 &  -7.43 &  -7.40 &   0.28 &   0.24 &   0.97 &   8.30 &   8.59 &   1.82 &   0.80 &   1.82 \\
 254 & 103 &  -7.43 &  -7.40 &   0.28 &   0.24 &   0.97 &        &        &        &        &        \\ \hline
 257 & 102 &  -7.43 &  -7.41 &   0.28 &   0.24 &   0.97 &   7.59 &   8.47 &   3.93 &   0.72 &   1.18 \\
 255 & 102 &  -7.44 &  -7.42 &   0.28 &   0.24 &   0.97 &   7.88 &   8.06 &   2.88 &   2.27 &   1.57 \\
 253 & 102 &  -7.45 &  -7.42 &   0.29 &   0.24 &   0.97 &   8.05 &   8.14 &   2.37 &   1.99 &   1.63 \\\hline
 252 & 101 &  -7.46 &  -7.44 &   0.28 &   0.24 &   0.97 &        &        &        &        &        \\\hline
 251 & 100 &  -7.48 &  -7.46 &   0.29 &   0.24 &   0.97 &        &        &        &        &        \\
 253 & 100 &  -7.47 &  -7.45 &   0.28 &   0.24 &   0.97 &        &        &        &        &        \\
 249 & 100 &  -7.49 &  -7.46 &   0.29 &   0.23 &   0.97 &        &        &        &        &        \\ \hline
\end{supertabular}
\end{center}
\end{small}

\begin{figure}[htb]
\centerline{\epsfig{file=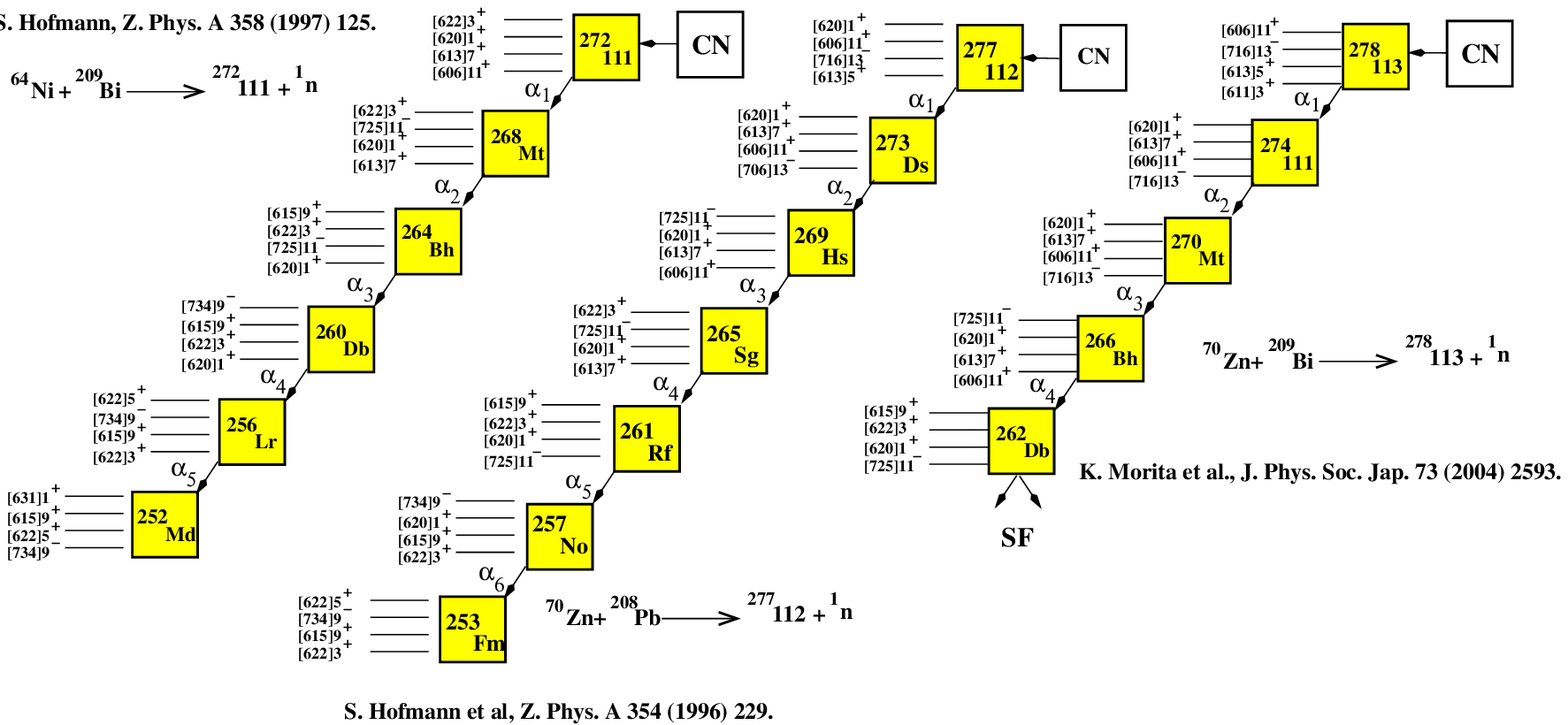,width=1.1\textwidth}}
\vspace{3cm}
\centerline{\epsfig{file=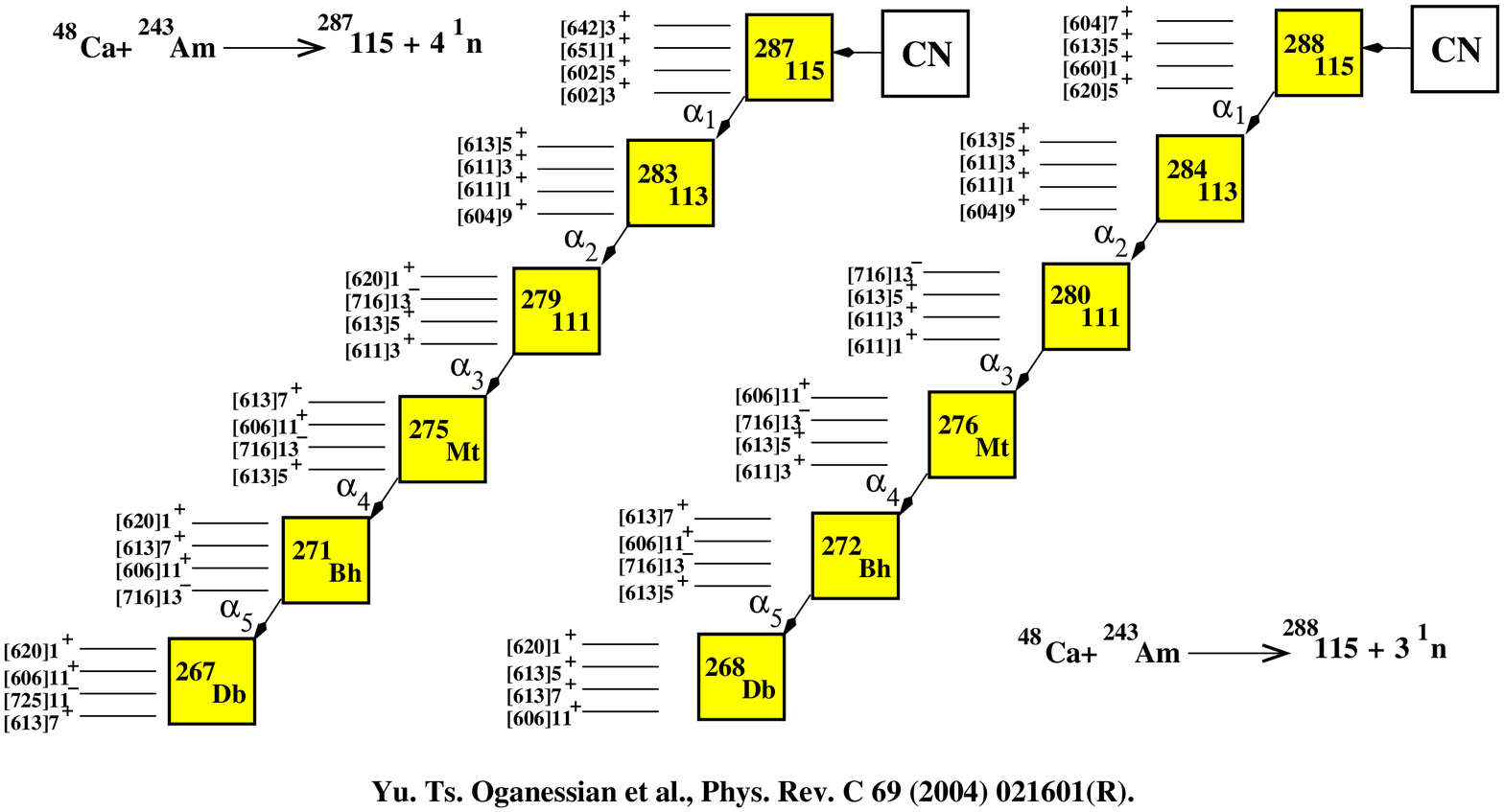,width=0.9\textwidth}}
\caption{Single particle level structure near the Fermi level}
\end{figure}

\subsubsection{$Q$ Values}

\begin{figure}[htb]
\centerline{\epsfig{file=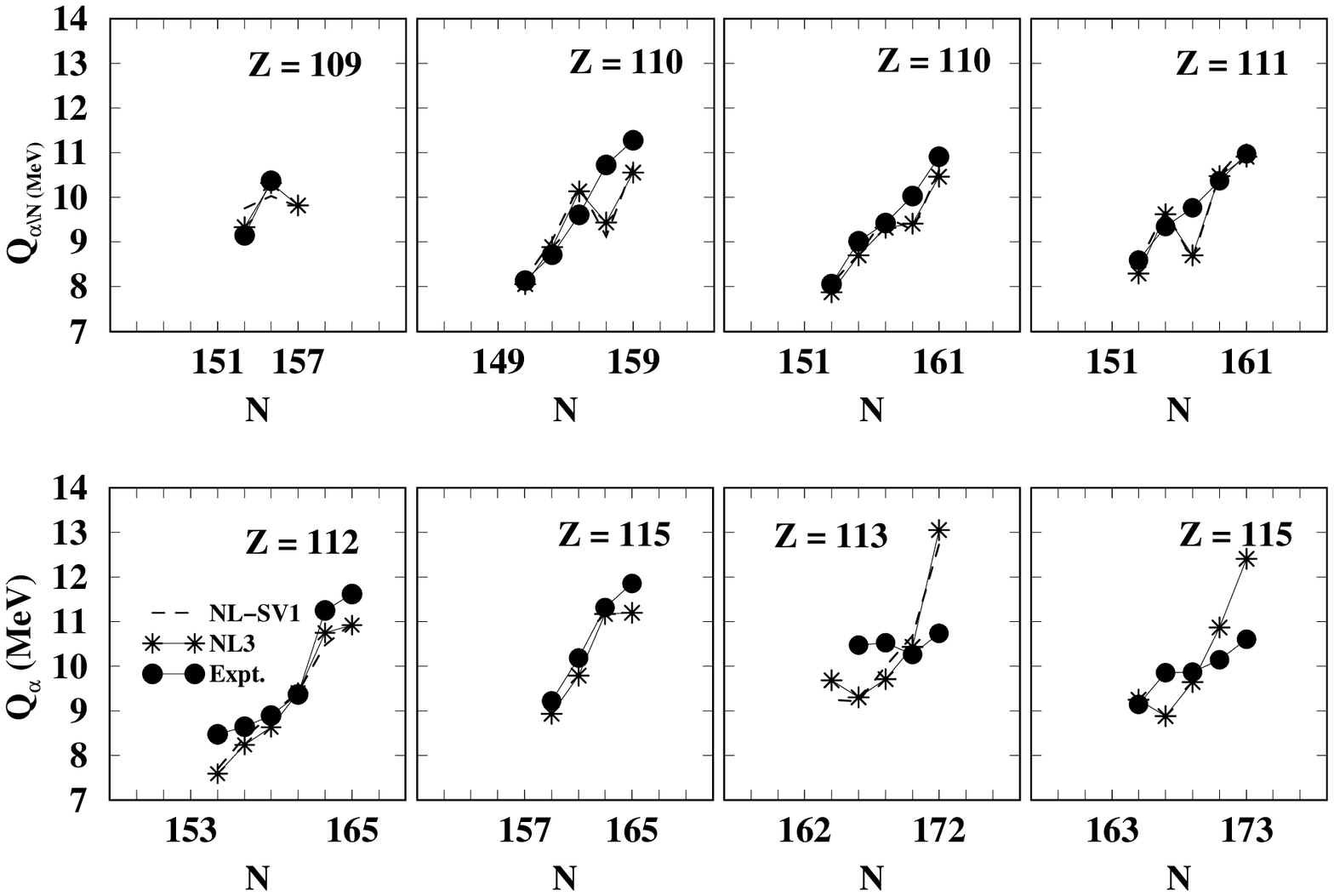,width=0.8\textwidth}}
\centerline{\epsfig{file=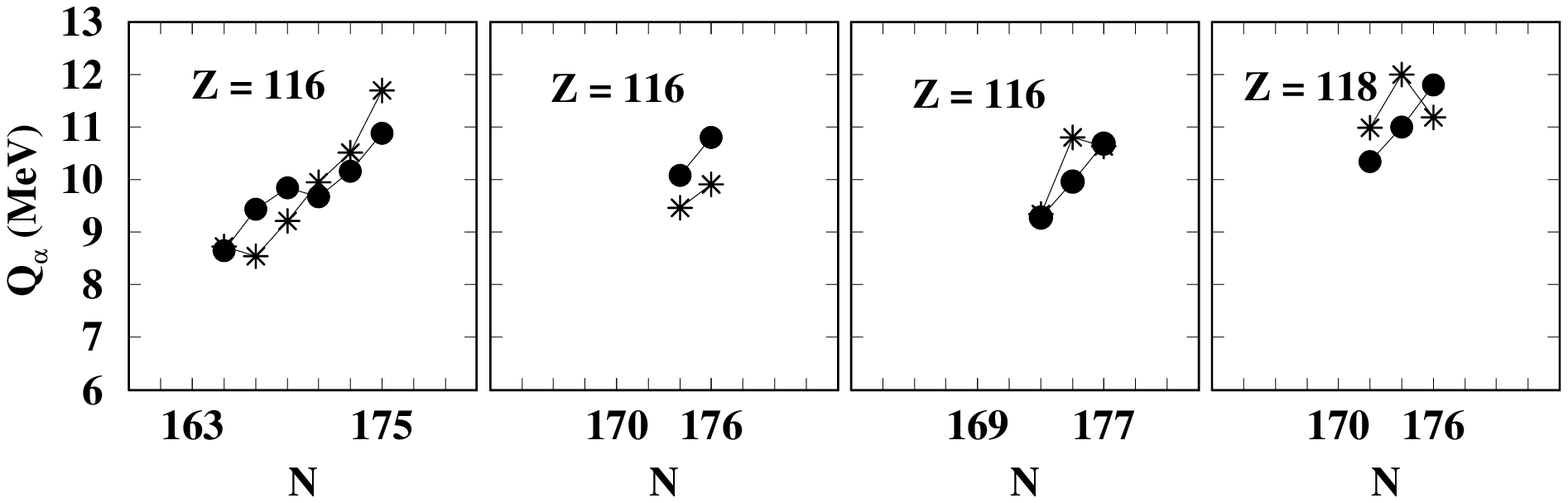,width=0.8\textwidth}}
\caption{The $Q$ values against $\alpha$ - decay. The corresponding experimental
values are also shown, where available.}
\label{qval}
\end{figure}
The $Q$ value of a parent nucleus against $\alpha$ - decay is simply the 
difference
between the binding energy of the parent nucleus and the sum of the 
binding energies
of both the resulting daughter and $\alpha$ particle. Here, we present Q-values
for the NL3 and NL-SV1 parameter sets in Fig. (\ref{qval}). 
Both calculations are found to be 
in agreement with 
each other, and also with the corresponding experimental values. The 
calculations deviate from experiment. At some places, the maximum departure 
is of the order of about 1 MeV, except for the decay chain on Z = 115, where
the deviation is about 2 MeV in a few cases. 
This level of agreement in indeed gratifying in view of the fact that 
the $Q$ value is the difference 
between large numbers. A small error in even one of them, could affect the 
$Q$ value substantially. 

\subsubsection{Matter Radii}

\begin{figure}[htb]
\centerline{\epsfig{file=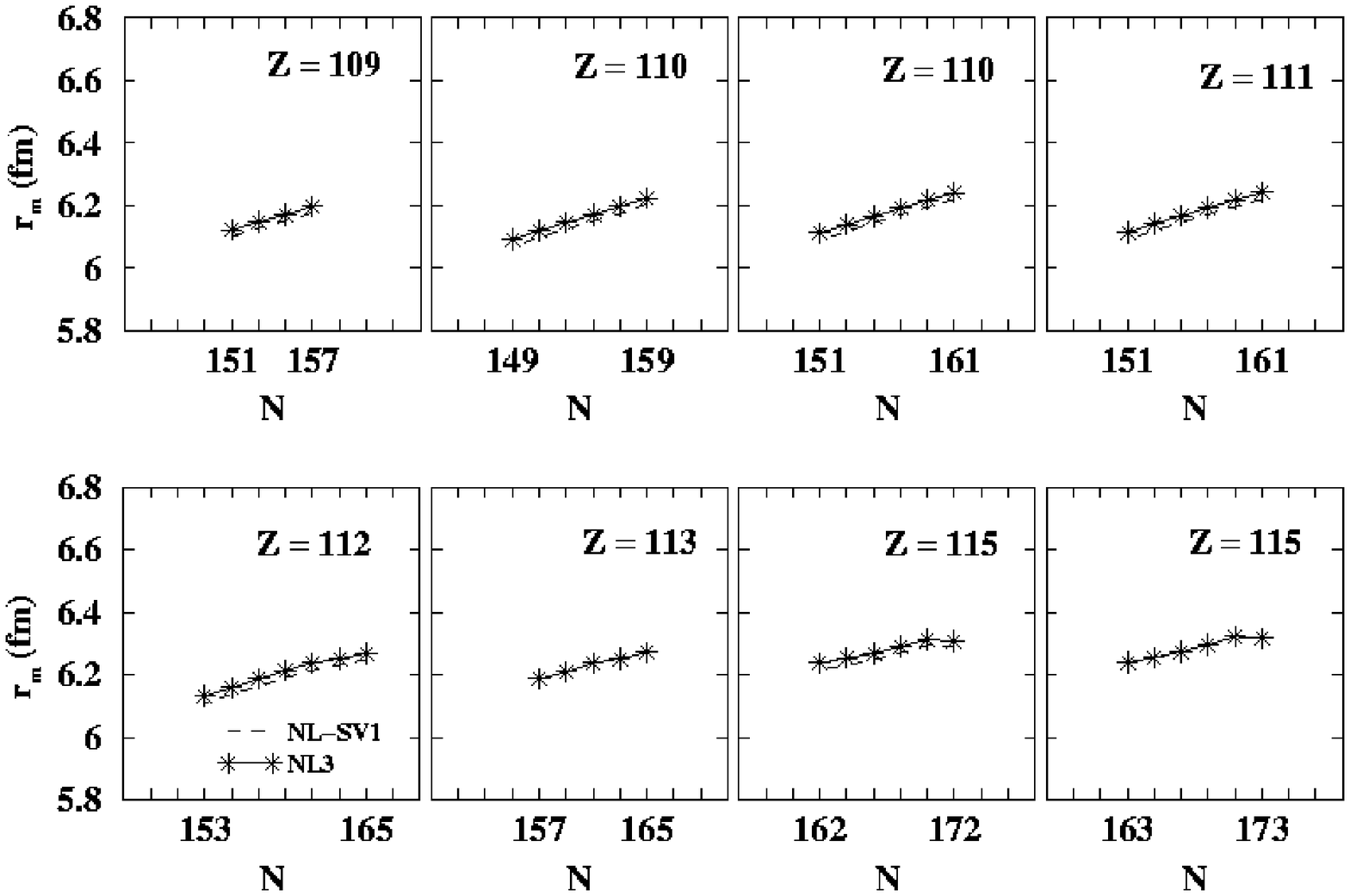,width=0.8\textwidth}}
\centerline{\epsfig{file=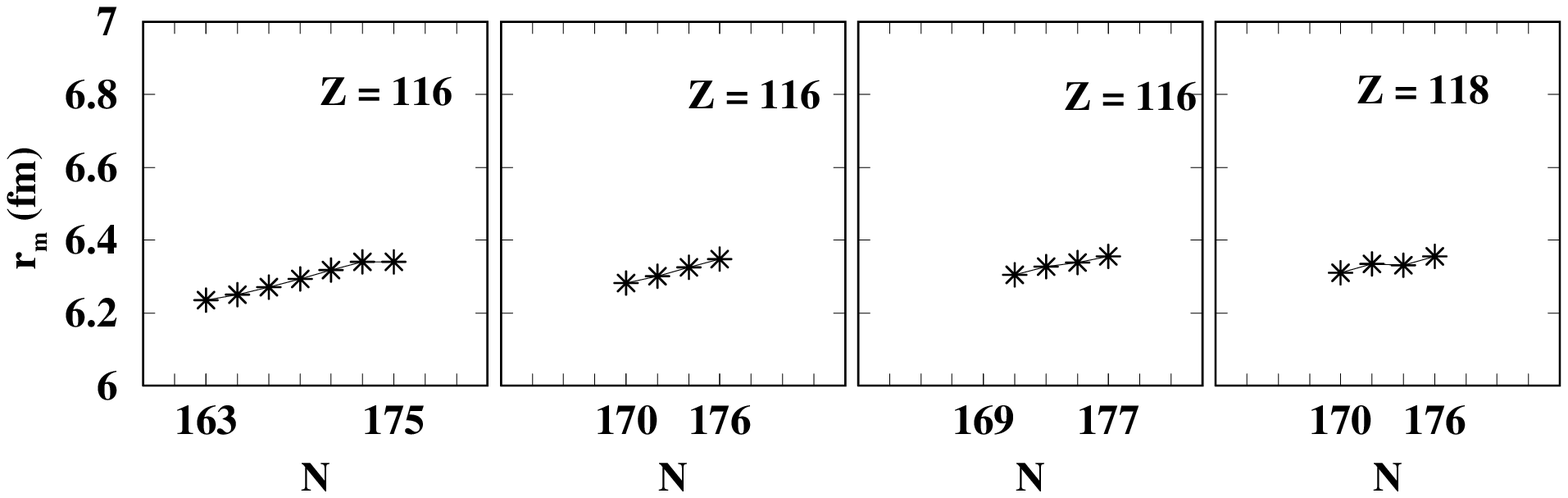,width=0.8\textwidth}}
\caption{The {\it rms} matter radii.}
\label{rm}
\end{figure}

\begin{figure}[htb]
\centerline{\epsfig{file=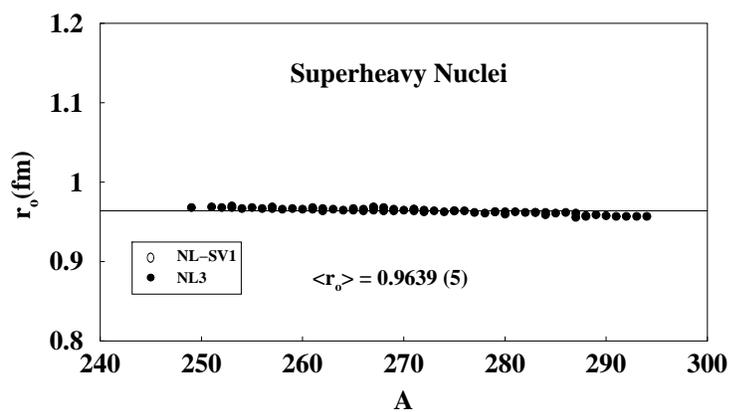,width=0.6\textwidth}}
\caption{The extracted $r_o$.}
\label{r_o}
\end{figure}

The root mean squared ({\it rms}) matter radii ($r_m$) are obtained from the 
{\it rms} proton and neutron radii through:
\begin{eqnarray} 
r_m^2~=~\frac{Zr_p^2~+~Nr_n^2}{Z+N}~.
\end{eqnarray} 
The calculated
$r_m$ values are shown in Fig. (\ref{rm}). The $r_m$ values are found to be
varying monotonically with mass number. 
The radius parameter $r_o$ extracted using the conventional relation:
\begin{eqnarray} 
r_m~=~r_oA^{1/3}
\end{eqnarray} 
is found to be a constant ($r_o~=~0.9639\pm0.0005$) for all the nuclei appearing
in the $\alpha$ - decay chains of SHE, shown in figure \ref{r_o}.

\clearpage

\subsection{Half Lives}

\subsubsection{Double Folding Model and WKB analysis}

The basic idea behind the double folding model is to get a reasonable 
nucleus-nucleus potential, knowing some nucleon-nucleon interaction (e.g. M3Y 
interaction, etc.). In general, the double folding potential comprises of the 
direct and the exchange terms. The direct term contains the direct nucleon - 
nucleon matrix elements, whereas, the exchange term, as the name suggests, 
contains the exchange part. The latter is considerably more difficult to handle 
in practice. Thus, for some of the applications, the exchange term is {\it 
simulated} by a delta function pseudo-potential, with some density dependence.

\begin{figure}[htb]
\centerline{\epsfig{figure=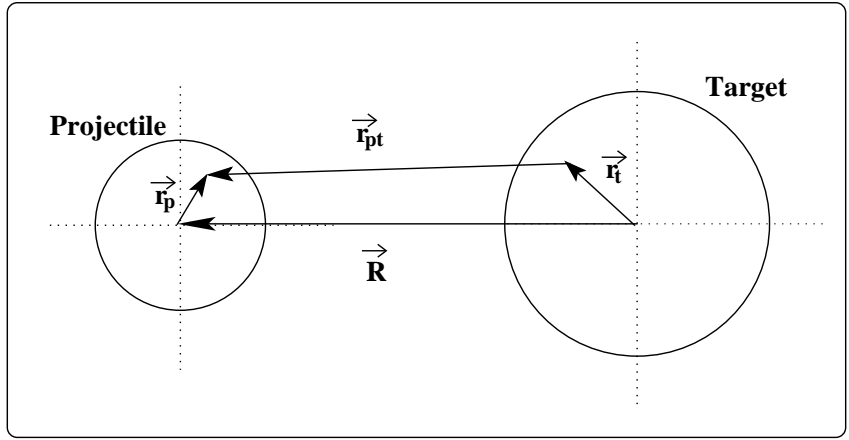,width=0.7\textwidth}}
\caption{Geometry of the Double Folding prescription.}
\label{DF_fig}
\end{figure}

In the present work, the double folding prescription to obtain the nucleus - 
nucleus potential has been used (refer to Fig. (\ref{DF_fig}) for the geometry of the problem). 
The M3Y effective nucleon - nucleon interaction employed here 
is given by \cite{SAT.79,AKC.86,KHO.94}:
\begin{eqnarray}
v^{M3Y}~=~7999\frac{e^{-4s}}{4s}~-~2134\frac{e^{-2.5s}}{2.5s}
\end{eqnarray}
The exchange effects are considered only through a delta function 
pseudo-potential \cite{AKC.86}:
\begin{eqnarray}
v^{pseudo} ~=~J_{00}(E)\delta (s)~;
\end{eqnarray}
where, the volume integral ($J_{00}$) is \cite{AKC.86}:
\begin{eqnarray}
J_{00}~=~-276(1~-~\frac{0.005E}{A_{\alpha}}) ~\mathrm{in~MeV~-~fm}^{3}
\end{eqnarray}
In the present work, the energy dependence is ignored. Thus, the M3Y interaction
with pseudo-potential becomes:
\begin{eqnarray}
v^{M3Y+pseudo}~=~7999\frac{e^{-4s}}{4s}~-~2134\frac{e^{-2.5s}}{2.5s}~-~276\delta (s) 
~~~~~~~\mathrm{in~MeV}
\end{eqnarray}
The density dependence is supposed to compensate to some extent the higher order
exchange effects and the effects of the Pauli Blocking. Following the earlier 
work \cite{AKC.86}, it is assumed to be of the form:
\begin{eqnarray}
v^{dd}~=~C(1~-~\beta (E)\rho_{1}^{2/3})(1~-~\beta (E)\rho_{2}^{2/3})
\end{eqnarray}
where, C is the overall normalization constant, and is taken to be 1.0 in the 
present work. $\beta (E)$ is the energy dependent part of the density dependent
term and is assumed to have a constant value, 1.6 \cite{AKC.86}. 

With the density dependence and the M3Y with pseudo-potential, the assumed 
nucleon-nucleon interaction is now given by:
\begin{eqnarray}
v(s)=C\left(7999\frac{e^{-4s}}{4s}-2134\frac{e^{-2.5s}}{2.5s}-276\delta (s)
\right) (1-\beta (E)\rho_{1}^{2/3})(1-\beta (E)\rho_{2}^{2/3})
\label{vint_psu}
\end{eqnarray}
In these expressions, $\beta (E)$ is 1.6 and $s$ is equal to $\vec{r}_{pt}$ 
(refer to the figure).

The total double folding potential between the nucleus - nucleus system is:
\begin{eqnarray}
V_{PT}(\vec{R}) = \int \rho_P(\vec{r}_p)\rho_T(\vec{r}_t)v(\vec{r}_p-\vec{r}_t+
\vec{R})
                  d\vec{r}_p d\vec{r}_t
\label{dfpot}
\end{eqnarray}
where, $v$ is as defined above (Eq. (\ref{vint_psu})). In the actual calculations, we evaluate the six 
dimensional integral in the above equation by transforming it into the momentum 
space. The details are easy to work out and are straightforward.

In order to evaluate the half life time of the nucleus against $\alpha$ - decay,
one needs to know the nuclear potential, the Coulomb potential and the energy 
of the alpha particle, which in turn is obtained by the $Q$ value and the zero point 
energy of the oscillation of the alpha particle in the potential well. The 
point to be 
noted here is, in the present case, it is assumed that the alpha particle is 
already 
formed in the parent nucleus, so that, its motion can be simulated by assuming 
that it is moving in an average potential well formed by the {\it daughter} - 
$\alpha$ system. The Coulomb potential required in this case, is obtained
by folding the point proton densities of target and projectile with the 
Coulomb interaction.

Here the decay half life is calculated using the WKB approximation. The half 
life time of the nucleus (parent) against the $\alpha$ - decay is then given by:
\begin{eqnarray}
T_{1/2}~=~\frac{ln 2}{\nu} \left(1~+~e^{K}\right)
\label{wkb_life}
\end{eqnarray}
where, within the WKB approximation, the action integral is given by:
\begin{eqnarray}
K~=~\frac{2}{\hbar}\int_{R_a}^{R_b} \left\{2\mu\left(E(R)-Q\right)
\right\}^{1/2}dR~,
\end{eqnarray}
and 
$$E(R) = V_{PT}(R) + V_C(R) + \frac{\hbar^2}{2\mu}\frac{\lambda^2}{r^2}$$
The third term represents the centrifugal barrier with $\lambda~=~l + 1/2$,
$l$ being the orbital angular momentum. 
The $R_a$ and $R_b$ are the lower and upper turning points respectively. These 
are determined through the requirement:
$$E(R_a) = Q + E_{\nu} = E(R_b)$$

In Eq. (\ref{wkb_life}), $\nu$ is the assault frequency, given by:
\begin{eqnarray} 
\nu ~=~\left(\frac{1}{2R}\sqrt{\frac{2E}{M}}\right)~,
\label{orig_nu}
\end{eqnarray} 
where, R is the radius of the parent, given by $R=1.2A^{1/3}$; $E$ is the 
energy of the alpha particle, corrected for recoil; $M$ is mass of alpha 
particle, expressed in MeV. 

We now present and discuss the calculated half live values.

\subsubsection{Half Lives}

\begin{figure}[htb]
\centerline{\epsfig{file=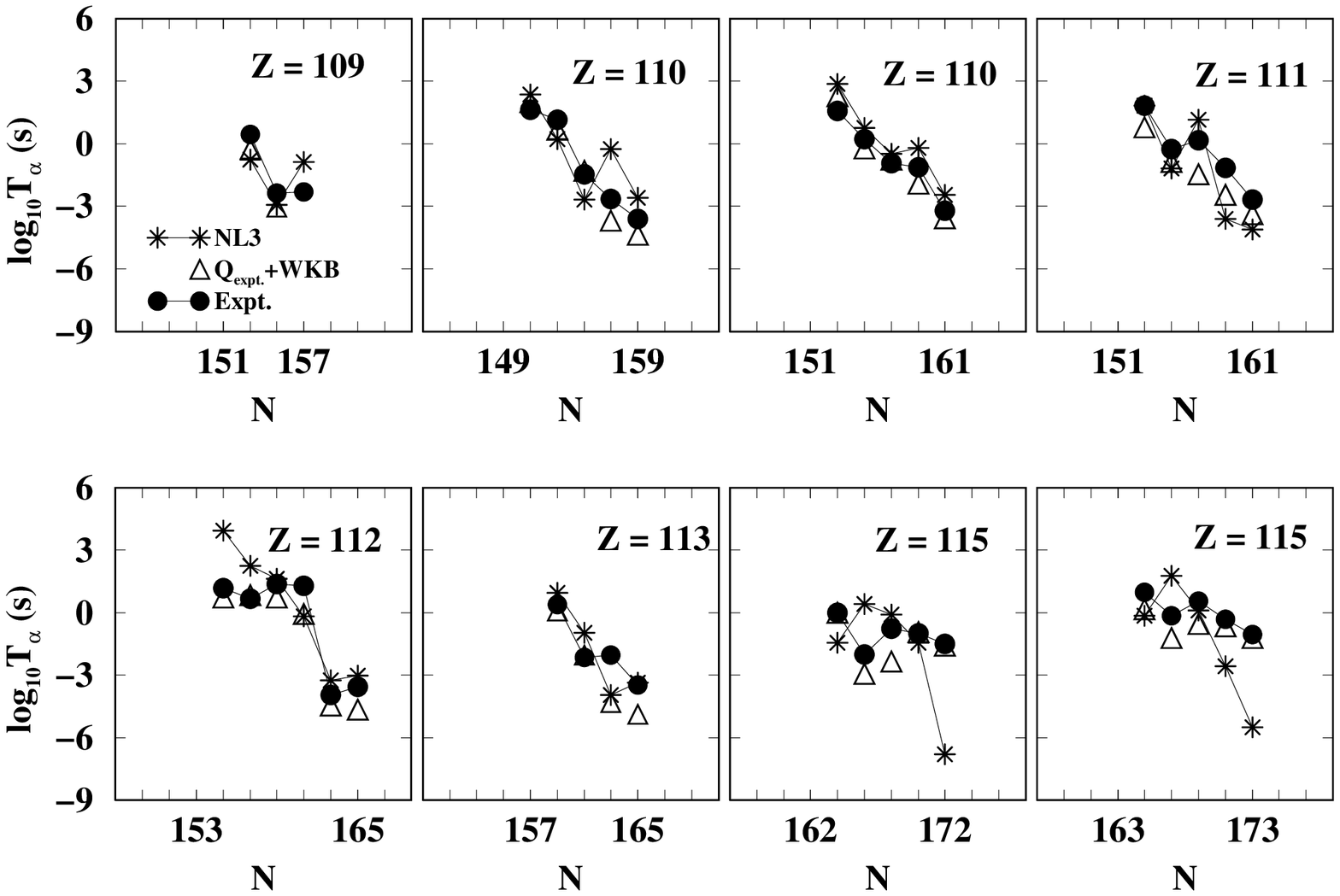,width=0.8\textwidth}}
\centerline{\epsfig{file=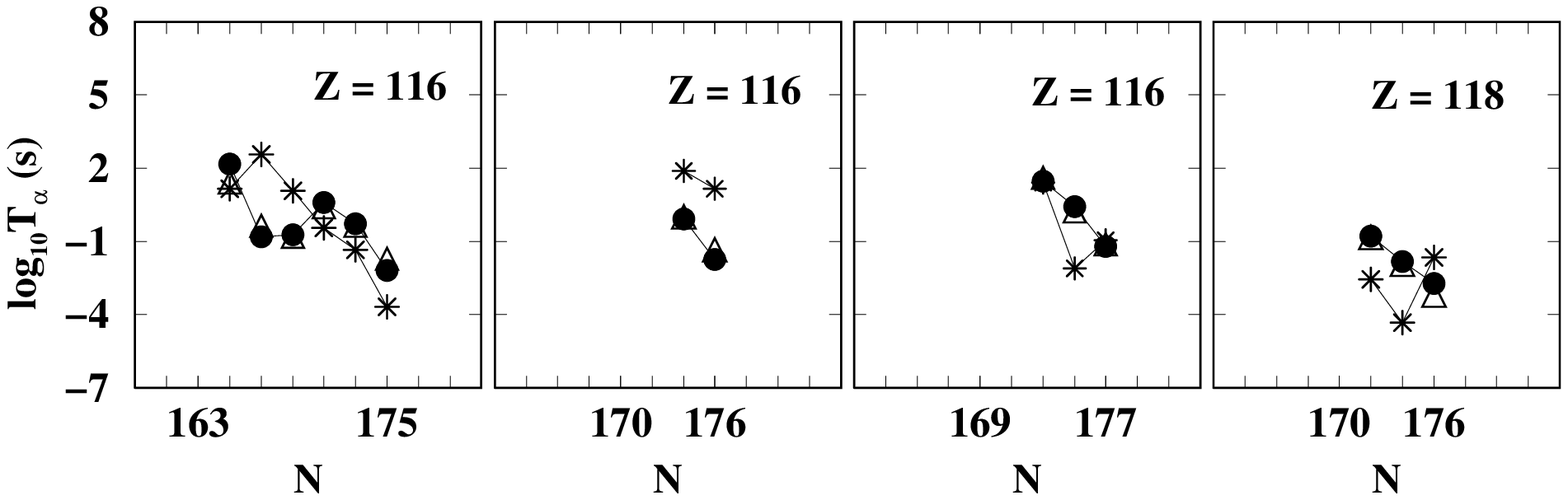,width=0.8\textwidth}}
\caption{Half lives against $\alpha$ - decay of superheavy nuclei.}
\label{thalf}
\end{figure}

The calculated and the corresponding experimental half lives for the superheavy
nuclei against $\alpha$ - decay are presented in Fig. (\ref{thalf}). Here, we 
have not presented the results for NL-SV1, since, we have already shown that 
the difference between NL3 and NL-SV1 results is negligible. 

Results of two independent calculations have been presented in Fig. 
(\ref{thalf}). In the first case, the calculated $Q$ values are used in the 
WKB procedure. These results are denoted by NL3. In the second case, the 
experimental $Q$ values are used in the WKB procedure, with the same microscopic
potential, as in the first case. These results are denoted by $Q_{expt.}+WKB$ 
in Fig. (\ref{thalf}). It is clearly seen that the use of experimental $Q$ 
values in WKB approach reproduces the experimental values rather well, 
indicating the reliability of the microscopic nucleus - nucleus potential, and
also validating the procedure that we follow in this work. The half lives 
obtained by using the calculated $Q$ values, though similar to the experimental
trend, differs from it quantitatively at places. This reflects hyper - 
sensitivity of the half lives on $Q$ values.

\section{Summary and conclusions}

Extensive and systematic microscopic self consistent RMF calculations for the nuclei 
appearing in the $\alpha$ - decay chains of the known superheavy elements are 
presented. The RMF results with only a few (seven) fixed Lagrangian parameters 
are found to reproduce the corresponding experimental ground state properties 
remarkably well. This indeed, is gratifying. The predicted neutron shell 
closures N $\sim$ 164, 172 and 184 are consistent with the recent experimental 
observations. The calculated root mean square radii closely follow the 
$A^{1/3}$ law. 

The interaction potential between the $\alpha$ and daughter nucleus,
obtained in the $t\rho\rho$ approximation using the RMF densities and the 
nucleon - nucleon interaction M3Y is employed to estimate the $\alpha$ - decay 
half lives. The calculations qualitatively agree with the experiment. The use 
of experimental $Q$ values brings the calculations closer to the experiment. 
Therefore, the calculated interaction potential 
is reliable and can be used with confidence in other reaction studies. 
 
\noindent 
{\bf Acknowledgments}

The authors are thankful to P. Ring, M. M. Sharma, J. Meng, 
G. M\"unzenberg, S. Hoffmann, K. Morita, Yu. Ts. Oganessian, M. G. Itkis,
E. M. Kozulin, E. Cherepanov, A. Sobiczewski, S. S. Kapoor and S. H. Patil 
for their  interest in this work. 
Partial financial 
support from the Board for Research in Nuclear Sciences (BRNS), Govt. of India 
(Proj. No. 2001/37/13/BRNS/485) is gratefully acknowledged. MG is thankful to 
Department of Science and Technology 
(DST), Govt. of India (Proj. No. SR/WOS-A/PS-108/2003) for financial support.


\end{document}